\def\year{2020}\relax
\documentclass[letterpaper]{article} 
\usepackage{aaai20}  
\usepackage{times}
\usepackage{helvet}
\usepackage{courier}
\usepackage{amsthm}
\usepackage{booktabs} 
\usepackage{amsmath,amssymb,amsfonts,chemarrow,balance}
\usepackage{hyperref}
\usepackage{pgfplots}
\usepackage{pst-plot}
\usepackage{lineno, blindtext}
\usepackage{amsmath} 
\usepackage{graphicx}
\usepackage{subfigure}
\usepackage{multirow}
\usepackage{diagbox}
\usepackage{balance}
\usepackage{mathenv}
\usepackage{breqn}
\newcommand{\BfPara}[1]{{\noindent\bf#1.}\xspace\xspace}
\usepackage{url}
\usepackage{cleveref}
\usepackage[most]{tcolorbox}
\usepackage{array}
\usepackage{stfloats}

\usepackage{amsmath}
\newcommand*\circled[1]{\tikz[baseline=(char.base)]{%
            \node[shape=circle,fill=black,draw,text=white,inner sep=0.5pt] (char) {#1};}}
\usepackage{color}
\usepackage[linesnumbered,ruled,noend]{algorithm2e}
\usepackage{algpseudocode}
\usepackage{amsmath}
\usepackage{graphics}
\usepackage{amssymb}
\usepackage{bm}
\usepackage{rotating}
\usepackage{epsfig}
\newcommand{\tsref}[1]{\textsection\ref{#1}\xspace}
\usepackage{verbatim}
\SetKwRepeat{Do}{do}{while}%
\makeatletter

\newcommand{\ie}{{\em i.e.,}\xspace}

\usepackage[inline]{enumitem}

\def\equationautorefname~#1\null{(#1)\null}
\usepackage{tikz,pgf}
\usetikzlibrary{arrows,backgrounds,snakes,decorations}
\usetikzlibrary{trees}
\usepackage{verbatim}
\usepackage[utf8]{inputenc}
\usepackage[english]{babel}

\usepackage{listings}
\usepackage{xcolor}
\usetikzlibrary{shadows} 

\colorlet{punct}{red!90!black}
\definecolor{background}{HTML}{F9F9F9}
\definecolor{delim}{RGB}{80,200,176}
\colorlet{numb}{blue!100!black}

\newtcbtheorem{Heuristic}{Heuristic}{
  enhanced,
  sharp corners,
  attach boxed title to top left={
    yshifttext=-1mm
  },
  colback=white,
  colframe=blue!5!black,
  fonttitle=\bfseries,
  boxed title style={
    sharp corners,
    size=small,
    colback=blue!5!black,
    colframe=blue!5!black,
  } 
}{thm}

\lstdefinelanguage{json}{
    basicstyle=\sffamily\footnotesize,
    showstringspaces=false,
    breaklines=true,
    frame=lines,
    backgroundcolor=\color{background},
    literate=
     *{0}{{{\color{numb}0}}}{1}
      {1}{{{\color{numb}1}}}{1}
      {2}{{{\color{numb}2}}}{1}
      {3}{{{\color{numb}3}}}{1}
      {4}{{{\color{numb}4}}}{1}
      {5}{{{\color{numb}5}}}{1}
      {6}{{{\color{numb}6}}}{1}
      {7}{{{\color{numb}7}}}{1}
      {8}{{{\color{numb}8}}}{1}
      {9}{{{\color{numb}9}}}{1}
      {:}{{{\color{punct}{\textbf{:}}}}}{1}
      {,}{{{\color{punct}{\textbf{,}}}}}{1}
      {\{}{{{\color{punct}{\{}}}}{1}
      {\}}{{{\color{punct}{\}}}}}{1}
      {[}{{{\color{punct}{[}}}}{1}
      {]}{{{\color{punct}{]}}}}{1},
}

\hypersetup{
	colorlinks,
	urlcolor=red!60!black,
	linkcolor=red!40!black,
	citecolor=blue!30!black,
	bookmarksnumbered
}

\newcommand{\RNum}[1]{\uppercase\expandafter{\romannumeral #1\relax}}

\begin{document}
%
\title{Towards Characterizing COVID-19 Awareness on Twitter}
\author{Muhammad Saad,\textsuperscript{1}
Muhammad Hassan,\textsuperscript{2}
Fareed Zaffar,\textsuperscript{3}\\
\textsuperscript{1}{University of Central Florida}\\
\textsuperscript{2}{University of Illinois Chicago}\\
\textsuperscript{2}{Lahore University of Management Sciences}\\
saad.ucf@Knights.ucf.edu,
mhassa42@uic.edu, 
fareed.zaffar@lums.edu.pk}
\maketitle
\begin{abstract}
The coronavirus (COVID-19) pandemic has significantly altered our lifestyles as we resort to minimize the spread through preventive measures such as social distancing and quarantine. An increasingly worrying aspect is the gap between the exponential disease spread and the delay in adopting preventive measures. This gap is attributed to the lack of awareness about the disease and its preventive measures. Nowadays, social media platforms (\ie Twitter) are frequently used to create awareness about major events, including COVID-19. In this paper, we use Twitter to characterize public awareness regarding COVID-19 by analyzing the information flow in the most affected countries. Towards that, we collect more than 46K trends and 622 Million tweets from the top twenty most affected countries to examine 1) the temporal evolution of COVID-19 related trends, 2) the volume of tweets and recurring topics in those trends, and 3) the user sentiment towards preventive measures. Our results show that countries with a lower pandemic spread generated a higher volume of trends and tweets to expedite the information flow and contribute to public awareness. We also observed that in those countries, the COVID-19 related trends were generated before the sharp increase in the number of cases, indicating a preemptive attempt to notify users about the potential threat. Finally, we noticed that in countries with a lower spread, users had a positive sentiment towards COVID-19 preventive measures. Our measurements and analysis show that effective social media usage can influence public behavior, which can be leveraged to better combat future pandemics. 

\end{abstract}

\section{Introduction and Related Work}\label{sec:intro}
The coronavirus (COVID-19) pandemic has spread across the world with over four million reported cases to date. Currently, no vaccine is available for the SARS-CoV-2 strain, and therefore the optimal way to curtail its spread is to avoid physical contact with COVID-19 carriers. To minimize the physical contact, people are advised to practice social distancing, stay at home, and in the worst case, undergo a lockdown~\cite{BroniecARG20,InoueT20}. Unfortunately, despite these guidelines, COVID-19 has spread faster than the adoption of preventive measures. The gap between the spread and the adoption of preventive measures is due to 1) limited awareness about the disease and its spread, 2) the nature of the disease and its latent symptoms~\cite{Robson20a}, and 3) delayed response in taking corrective measures by governments and the general public. Particularly, the aspect of public awareness largely depends on the information spread through the mainstream media and the social media~\cite{WellsSLPPY20,LeSS17,Brena0CGPR19}. Between these two axes of communication, social media platforms (\ie Twitter and Facebook) are highly useful in propagating timely information regarding a major event~\cite{TareafBHKKMC18}. Therefore, it is intuitive to assume that social media platforms contain information footprints that can be leveraged to characterize the response of various communities to the COVID-19 pandemic. To that end, this study uses Twitter data to analyze various attributes of information exchange in order to model preparations of various countries for the COVID-19 pandemic.

For this study, we draw inspiration from prior related works that have demonstrated the usefulness of Twitter in characterizing the user behavior in major events. For instance, \cite{AnYLDZL18} showed that during the Ebola pandemic, Twitter users actively discussed the risk potential and the spread rate. Similarly~\cite{Fischer-Pressler19}  and~\cite{KeymaneshGBBCP19} showed that Twitter is useful in monitoring the {\em social efficacy} and collective understanding of masses during critical global events. We follow a similar methodology and use Twitter to study the response of various countries to the COVID-19 pandemic. We collect trends and tweets from the top 20 most affected countries by COVID-19 (as of April 19, 2020) and contextualize the information to study their preparatory response. More precisely, using our dataset, we explore the following key questions. 
\begin{enumerate}
    \item {\em Are there variations in the response of different countries to the COVID-19 pandemic that are reflected in rends and tweets from that country?}    
    \item {\em Are there indications to support that awareness through Twitter was useful in influencing the pandemic spread?}
\end{enumerate}

\begin{figure*}[t]
    \centering
  \includegraphics[ width=1\textwidth]{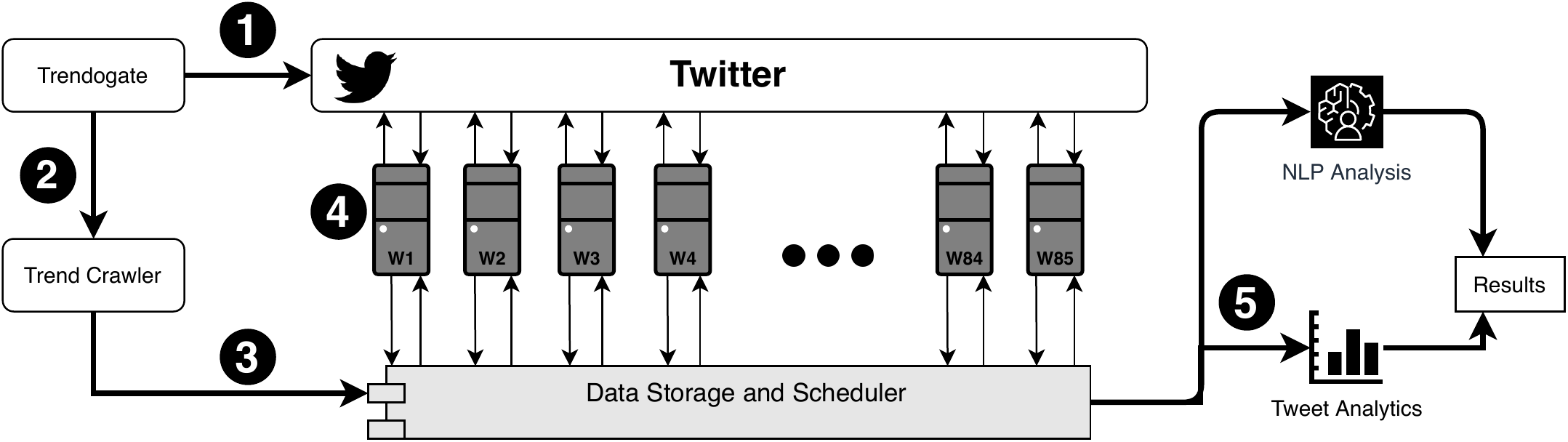}
  \vspace{-2mm}
    \caption{Design and workflow of our data collection system. First, we deployed a crawler to collect trends from {\em Trendogate}, and store them in the ``Data Storage and Scheduler'' platform hosted on cloud. The scheduler fed search queries to eighty-five workers that collected tweets from trends. Finally, we applied data analytics and NLP to collect results.    }
    \label{fig:dc}
\vspace{-3mm}
\end{figure*}

In pursuit of these questions, we develop a data collection system to collect more than 5,000 Twitter trends and over 622 Million tweets from the top 20 most affected countries by COVID-19 (as of April 19, 2020). For each country, we monitor the temporal patterns of COVID-19 trends and the volume of tweets in those trends to study the country's response to the pandemic. We perform topic modeling and sentiment analysis on tweets to analyze the user response towards preventive measures such as social distancing, quarantine, and lockdown. Our dataset reveals meaningful insights, including a correlation between frequent trending on COVID-19 and effective pandemic management. To illustrate this observation, we provide a comparative case study of six countries (USA, Italy, Spain, Sweden, Austria, and Belgium), which indicates that countries with a lower pandemic spread usually generated more tweets and trends about COVID-19 and its preventive measures. We believe that the key takeaways of our work highlight the importance of social media in influencing public interactions that can be useful in combating future pandemics.

\BfPara{Contributions and Roadmap} We take a systematic approach towards analyzing the temporal evolution of Twitter trends in 20 most affected countries by COVID-19. Our data collection, methodology, and results are summarized below as the key contributions.

\begin{enumerate*}[label=\protect\circled{\arabic*}, font=\small\bfseries]
    \item We developed a data collection system using which we collect more than 48K trends and over 622 Million tweets from December 15, 2019, to April 5, 2020, for the top 20 countries affected by COVID-19.

    \item For each country, we identify the COVID-19 related trends among all trends and the volume of tweets in those trends. We pair that information with key indicators in the country's COVID-19 timeline to study their preparatory status. 
    
    \item We present a case study of six countries (United States, Italy, Spain, Sweden, Austria, and Belgium) to a) show the variation in response of each country to the pandemic, and b) showcase observations that suggest that frequent and timely information propagation about preventive measures correlated with a lower pandemic spread. Notably, our results show that on average, Sweden, Austria, and Belgium generated more trends and tweets about COVID-19 and its preventive measures than the United States, Italy, and Spain. 
    
    \item Additionally, we apply Natural Language Processing (NLP) to extract the most prevalent topics in the COVID-19 tweets and the user sentiment towards those topics. We observed that countries with a lower pandemic spread frequently used terms related to the preventive measures such as ``social distancing.'' 
\end{enumerate*}

The rest of the paper includes data collection and methodology in~\tsref{sec:dc}, experiments and results in~\tsref{sec:exp}, discussion in~\tsref{sec:conc}, and appendices with supplementary findings in~\tsref{sec:append}. 
\section{Data Collection and Methodology} \label{sec:dc}   

This section describes our data collection and methodology. We started data collection on April 19, 2020, by selecting the top 20 most affected countries on that date. For each country, we collected all their Twitter trends from December 15, 2019, to April 5, 2020, .~\autoref{fig:dc} shows our data collection system, and below, we briefly discuss some key challenges that we encountered during the process.

\begin{figure*}[t]
    \centering
  \includegraphics[ width=1\textwidth]{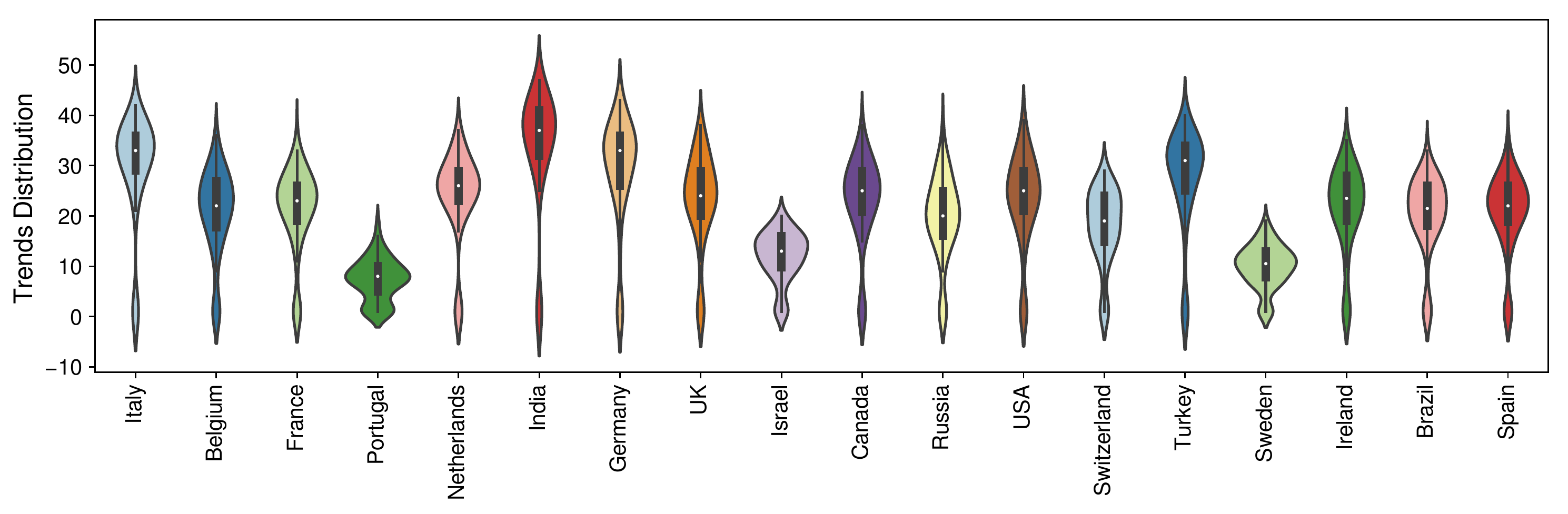}
  \vspace{-10mm}
    \caption{Distribution of trends shown as a violin plot for each country in our dataset. On average, India had the most trends, followed by Italy, Germany, and Turkey. In contrast, Portugal, Israel, and Sweden had the least number of trends. Violinplot shows the smooth distribution of data as a result. The result of the smoothing function can exceed the maximum and minimum value. As such, this should not be confused with the negative number of trends for the countries shown in the figure.}
    \label{fig:tr_one}
\vspace{-4mm}
\end{figure*}

\begin{table*}[!t]
\centering
\caption{Results from data collection. Each country is ranked based on the total number of COVID-19 cases as of April 19, 2020. Note that 1) India generated the highest trends, 2) Switzerland generated the highest COVID-19 trends, the highest overall tweets, 3) Ireland generated the highest tweets before the first case, 4) Belgium generated highest tweets before the first death,  5) Switzerland generated the highest COVID-19 tweets, and 6) Turkey generated the highest number of trends before the first case and the first death. Percentages are reported relative to the total number trends and the total number of tweets.}

\scalebox{0.72}{\begin{tabular}{lllllllll}
\hline\hline
\multicolumn{1}{l}{\textbf{Country}} & \multicolumn{1}{c}{\textbf{\begin{tabular}[c]{@{}c@{}}Total \\ Trends\end{tabular}}} & \multicolumn{1}{c}{\textbf{\begin{tabular}[c]{@{}c@{}}COVID-19\\  Trends\end{tabular}}} & \multicolumn{1}{c}{\textbf{\begin{tabular}[c]{@{}c@{}}Total \\ Tweets\end{tabular}}} & \multicolumn{1}{c}{\textbf{\begin{tabular}[c]{@{}c@{}}COVID-19\\ Tweets\end{tabular}}} &
\multicolumn{1}{c}{\textbf{\begin{tabular}[c]{@{}c@{}}COVID-19 Trends\\ before first Case\end{tabular}}} &
\multicolumn{1}{c}{\textbf{\begin{tabular}[c]{@{}c@{}}COVID-19 Tweets\\ before first Case\end{tabular}}} &
\multicolumn{1}{c}{\textbf{\begin{tabular}[c]{@{}c@{}}COVID-19 Trends \\before first Death\end{tabular}}} &
\multicolumn{1}{c}{\textbf{\begin{tabular}[c]{@{}c@{}}COVID-19 Tweets \\before first Death\end{tabular}}} \\
\hline\hline
USA & 2,553 & 54 (2.12\%) & 32,283,958 & 852,271 (2.64\%) & 3 (0.12\%) & 20,855 (0.06\%) & 12 (0.47\%) & 316,590 (0.98\%) \\
Spain & 2,841 & 90 (3.17\%) & 27,864,760 & 451,017 (1.62\%)& 14 (0.49\%) & 74,486 (0.27\%) & 15 (0.53\%) & 74,752 (0.27\%) \\
Italy & 3,244 & 94 (2.90\%) & 41,378,010 & 812,820 (1.96\%)& 13 (0.40\%) & 176,164 (0.43\%) & 23 (0.71\%) & 306,630 (0.74\%) \\
France & 2,337 & 34 (1.45\%) & 24,677,002 & 1,075,032 (4.36\%)& 7(0.30\%)  & 220,888 (0.90\%) & 14(0.60\%) & 392,492 (1.59\%) \\
Germany & 3,562 & 213 (5.98\%) & 43,815,322 & 2,094,764 (4.78\%)& 19 (0.53\%) & 252,958 (0.58\%) & 73 (2.05\%) & 944,062 (2.15\%) \\
UK & 2,814 & 92 (3.27\%) & 28,760,308 & 822,120 (2.86\%)& 26 (0.92\%) & 71,541 (0.25\%) & 47 (1.67\%) & 361,071 (1.26\%) \\
Turkey & 3,511 & 225 (6.41\%) & 24,974,434 & 924,090 (3.70\%) & 139 (3.96\%) & 827,429 (3.31\%) & 154 (4.39\%) & 864,636 (3.46\%) \\
Russia & 2,214 & 112 (5.06\%) & 37,572,533 & 3,434,981 (9.14\%)& 14 (0.63\%) & 459,865 (1.22\%) & 16 (0.72\%) & 511,323 (1.36\%) \\
Brazil & 2,267 & 27 (1.19\%) & 26,578,248 & 218,389 (0.82\%)& 10 (0.44\%) & 43,129 (0.16\%) & 20(0.88\%) & 214,664 (0.81\%) \\
Belgium & 1,973 & 218 (11.05\%) & 38,118,669  & 3,188,033 (8.36\%)& 24 (1.22\%) & 739,441 (1.94\%) & 59 (2.99\%) & 1,536,010 (4.03\%) \\
Canada & 2,586 & 98 (3.79\%) & 30,866,694 & 1,006,908 (3.26\%)& 18 (0.70\%) & 421,630 (1.36\%) & 38 (1.47\%) & 633,268 (2.05\%) \\
Netherlands & 2,707 & 205 (7.57\%) & 35,808,148 & 3,095,914 (8.65\%)& 46 (1.70\%) & 689,828 (1.93\%) & 66 (2.44\%) & 728,226 (2.03\%) \\
Switzerland & 2,055 & 272 (13.24\%) & 35,635,280 & 7,374,920 (20.70\%)& 56 (2.73\%) & 661,132 (1.86\%) & 65 (3.16\%) & 826344 (2.32\%) \\
Portugal & 827 & 78 (9.43\%) & 8,024,299 & 827,998 (10.32\%)& 31 (3.75\%) & 522,049 (6.51\%) & 38 (4.59\%) & 672,202 (8.38\%) \\
India & 3,746 & 85 (2.27\%) & 31,167,625 & 876,260 (2.81\%)& 18 (0.48\%) & 240,498 (0.77\%) & 36 (0.96\%) & 387,352 (1.24\%)\\ 
Peru & 1,980 & 74 (3.74\%)  & 31,479,593 & 1,066,531 (3.39\%) & 13 (0.66\%) & 582,652 (1.85\%)  & 39 (1.97\%) & 929,755 (2.95\%)  \\
Ireland & 2,739 & 224 (8.18\%) & 44,327,789 & 2,462,102 (5.55\%)& 75 (2.74\%) & 1,055,562 (2.38\%) & 93 (3.40\%) & 1,191,083 (2.69\%)\\
Austria & 1,523 & 131 (8.60\%) & 31,328,988 & 4,859,019 (15.51\%)& 26 (1.71\%) & 630,166 (2.01\%) & 46 (3.02\%) & 1,098,052 (3.50\%) \\
Sweden & 1,328 & 119 (8.96\%) & 18,173,812 & 1,986,659 (10.93\%)& 17 (1.28\%) & 481,854 (2.65\%) & 33(2.48\%) & 681,829 (3.75\%) \\
Israel & 1472 & 113 (7.68\%) & 30,141,353 & 5,500,852 (18.25\%)& 32 (2.17\%) & 837,435 (2.78\%) & 81 (5.50\%) & 4,281,735 (14.21\%) \\
\hline\hline
\textbf{Total} & 48,279 & 2,558 (5.30\%) & 622,976,825 & 42,930,680 (6.89\%)& 601 (1.24\%) & 9,009,562 (1.45\%) & 968 (2.01\%) & 16,952,076 (2.72\%)\\
\hline\hline           
\end{tabular}}\label{tab:one}
\end{table*}

\BfPara{Collecting Historical Trends} A trend on Twitter generally indicates a commonly discussed topic by users in a location~\cite{TulasiGGSMBK19}. Logically, on a specific date, if all trends of a location are collected, we can estimate the commonly discussed topics in that region. As such, the first challenge in our study was to obtain all the historical trends of the selected countries. Twitter API does not provide historical trends for countries, and therefore, we relied on third-party services for trend collection. We used an online service called {\em Trendogate} that maintains historical Twitter trends for all countries~\cite{trendogate20}. We developed a crawler that periodically scraped trends of each country and stored them in our ``Data Storage and Scheduler'' platform. For validation, we cross-examined those trends with an Internet archive service called ``Wayback Machine''~\cite{wayback_machine20}. The ``Wayback Machine'' creates snapshots of a vast majority of the Internet webpages every day. Currently, the archive contains historical data of more than 330 billion web pages. After cross-examining and validating data, our `Data Storage and Scheduler'' platform created a list of Trends for eighty-five workers that we deployed for tweet collection (\autoref{fig:dc}).

\BfPara{Collecting Tweets from Trends} We developed twitter crawlers and deployed them on eighty-five workers for concurrent data collection. We could not use the Twitter API since the API only provides the recent tweets from a trend. To overcome this limitation, we developed web crawlers that utilized Twitter's scroll loader functionality to collect tweets. Each web crawler generated a search query for a trend in a country and iterated over the scroll loader to scrape tweets. For this purpose, we sought help from prior works that have utilized Twitter's scroll loader functionality for data collection~\cite{Pratikakis18,mottl_2019,ValkanasSG14}. We also noticed that Twitter applies rate-limiting on IP addresses that generate iterative search queries. Keeping in view the desirable data volume, we provisioned 85 workers and replicated our crawler on these workers. Each worker was represented by a unique IP Address over the Internet. Upon receiving a rate-limiting error, each worker applied a linear backoff time. \autoref{fig:dc} provides the data collection system workflow. 

\subsection{Methodology and Preliminary Results} \autoref{tab:one} and~\autoref{fig:tr_one} show preliminary results obtained after data collection. At the time of the writing of this paper, we were able to collect trends and tweets from December 15, 2019, to April 5, 2020. Therefore, the results reported in this study are confined within this timeline.~\autoref{fig:tr_one}, reports the distribution of daily trends obtained from each country as violin plots. For each plot, the white dot in the middle is the median value of the total number of trends, the grey bar is the interquartile range, and the outer shape is the kernel density estimation showing the data distribution (details of kernel density function in~\autoref{sec:kd}).~\autoref{fig:tr_one} shows that 1) the number of daily trends from each country varied from as low as one trend in Israel to 47 trends in Italy, and 2) the average number of daily trends was 21. Therefore, we expected variations in the duration of data collection for each country, and our ``Data Storage and Scheduler'' platform applied load-balancing to maximize the system utility. 

\BfPara{Trend Collection} In~\autoref{tab:one}, we report the preliminary results in which we collect statistics about the first case and the first death from~\cite{COVID19Timeline}. For each country, we record the total number of 1) all trends, 2) all tweets, 3) trends related to COVID-19, 4) tweets related to COVID-19, 5) the number of trends and tweets before the first reported case, and 6) the number of trends and tweets before the first reported the death. All countries are sorted in the descending order as of April 19, 2020 (when we began our study), where the United States had the highest number of COVID-19 cases followed by Spain and Italy, respectively. To obtain the COVID-19 related trends and tweets, we curated a list of COVID-19 terms from the Yale Medicine Glossary~\cite{katella_2020} and Texas Medical Center~\cite{tmc_2020}. For each country, we matched the trend string and the tweet text with the set of COVID-19 terms to prepare~\autoref{tab:one}. For countries other than the United States, Canada, and the United Kingdom, we also translated the COVID-19 terms in their native language to maximize precision and obtain complete information about tweets and trends. Overall, we collected 48,279 trends from 20 countries with 2,558 (5.30\%) trends related to COVID-19. The highest number of total trends (3,746) were generated in India, and only 85 among them were related to COVID-19. The lowest number of trends (827) were generated from Portugal, and 78 among them were related to COVID-19. Switzerland created the highest number of COVID-19 trends (272). In contrast, France generated the minimum amount of COVID-19 trends (34). Turkey generated the highest number of COVID-19 trends before the first case and the first death (139 and 154, respectively).

\BfPara{Tweet Collection} From all these trends, we collected more than 622 Million tweets with 42 Million tweets related to COVID-19. Ireland generated the highest number of tweets ($\approx$44.3 Million) with $\approx$2.4 Million tweets related to COVID-19. Portugal generated the lowest number of tweets ($\approx$8 Million) with $\approx$82K tweets related to COVID-19. Switzerland generated the highest number of COVID-19 tweets ($\approx$ 7.3 Million among $\approx$35 Million 20.70\% tweets), while Brazil generated the lowest number of COVID-19 tweets ($\approx$218K among $\approx$26 Million tweets). Ireland generated the highest number of tweets before the first reported case ($\approx$1 Million), and the highest number of tweets before the first death ($\approx$ 1.1 Million). It is noteworthy in our data that no country, among the top three most affected, generated the maximum number of trends or tweets.

 \BfPara{Limitations} During data collection, we could not collect trends from China due to a state-backed ban on Twitter. Another limitation of our work is that {\em Trendogate} reported limited visibility into Iran. As a result, we could collect trends from Iran. We exclude these countries from our study, and since they were among the top 20 countries, we omitted them and added Ireland and India that were on 21st and 22nd positions at the time of this study. Despite these limitations, our dataset covers a wide range of countries that can sufficiently provide results required for our analysis.

\section{Experiments and Results} \label{sec:exp}
We conduct three experiments to analyze the response of each country to the COVID-19 pandemic. In the first experiment, we analyze the temporal behavior of COVID-19 related trends and tweets to study the patterns in the information spread. We present a case study of six countries to highlight the variation in response, characterized by the number of trends and the tweet volume in those trends. In the second and third experiments, we perform topic modeling and sentiment analysis to study the commonly discussed COVID-19 topics and the user sentiment in those discussions. 

\subsection{Temporal Analysis} \label{sec:temporal}
For temporal analysis, we specify a timeline for each country where we observe the total number of trends and tweets generated every day. Our timeline starts from December 21, 2019, to April 5, 2020. We exclude trends and tweets before December 21 since we did not observe any significant COVID-19 related data to report. We made the following key observations in the temporal analysis.

\setlength{\textfloatsep}{1pt}
\begin{algorithm}[t] 
\SetAlgoLined\SetArgSty{}
\textbf{Input:} List of COVID-19 terms \textsf{Terms}, List of Dates \textsf{Dates}, List of all Tweets \textsf{Tweets}, where an element \textsf{tweet} $\in$ \textsf{Tweets} is an object (\text{trends},\text{text}) \\
\textbf{Initialize:} TwList, TrList \\
\ForEach{\textsf{date} $\in$ \textsf{Dates} }{
\ForEach{\textsf{tweet} $\in$ \textsf{Tweets} }{
    \If{ \textsf{tweet}.\text{text} $\in$  \textsf{Terms} \textbf{and} \textsf{tweet}.\text{trends} $\in$  \textsf{Terms} }{ 
TwList $\leftarrow$\textsf{tweet}.\text{text} , TrList $\leftarrow$\textsf{tweet}.\text{trend}  \\}

    \If{ \textsf{tweet}.\text{trend} $\notin$  \textsf{Trends} \textbf{and} \textsf{tweet}.\text{text} $\in$ \textsf{Terms}  }{ 
    TwList $\leftarrow$\textsf{tweet}.\text{text} \\}
    
    \If{ \textsf{tweet}.\text{trend} $\in$  \textsf{Terms} \textbf{and} \textsf{tweet}.\text{text} $\notin$ \textsf{Terms}  }{ 
TwList $\leftarrow$\textsf{tweet}.\text{text} , TrList $\leftarrow$\textsf{text}.\text{trend}  \\}
 
}
}
\SetKwInput{KwData}{return} 
 \KwData{TwList for \autoref{fig:a}, TrList  for \autoref{fig:b}}  
\caption{Identifying COVID-19 trends and tweets in $\mathcal{S}_{1}$ and $\mathcal{S}_{2}$ for \autoref{fig:cstudy}}
  \label{algo:sk}
\end{algorithm}

\begin{figure*}[t]
    \centering
\subfigure[Total Number of tweets in COVID-19 related trends. Overall, Sweden, Austria, and Belgium produced more Tweets compared to the other three countries, indicating a higher user engagement. Austria produced the highest number of Trends and Tweets. The inner plot shows the total number of COVID-19 trends generated in a day. Belgium generated a maximum of 15 trends on March 19, 2020.]{\label{fig:a}\includegraphics[width=1\textwidth]{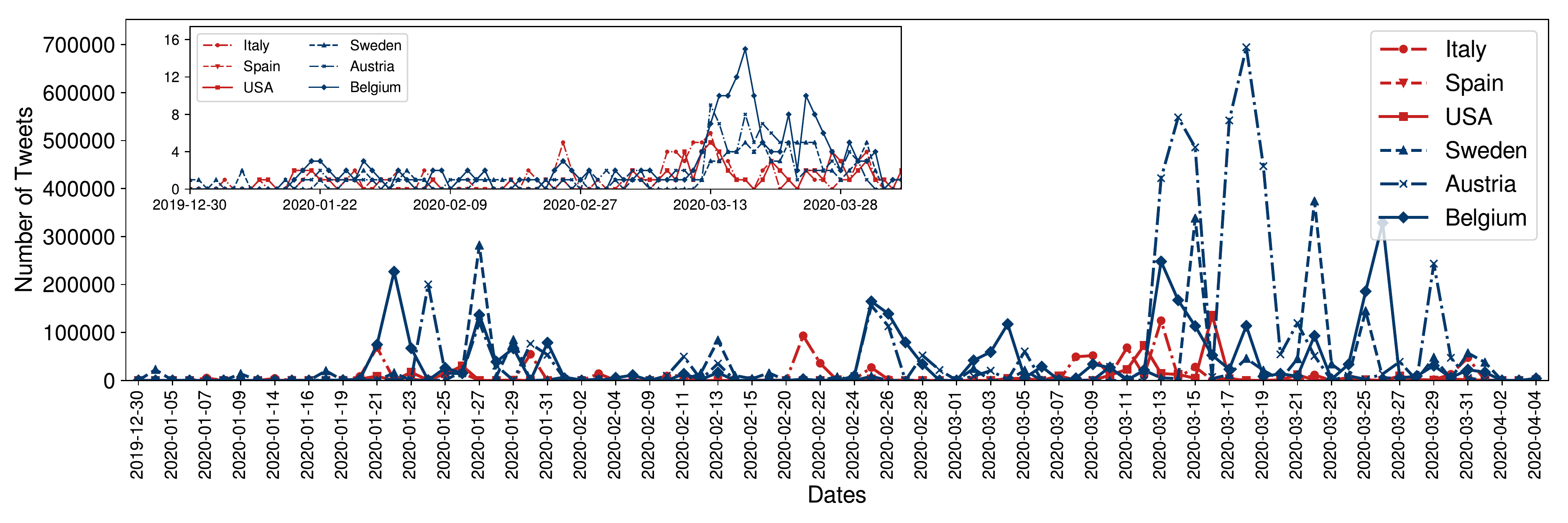}}

\subfigure[Total number of COVID-19 related tweets among all trends. Overall, the results are consistent with \autoref{fig:b}, showing that Sweden, Austria, and Belgium generated more COVID-19 tweets compared to the other three countries.   ]{\label{fig:b}\includegraphics[width=1\textwidth]{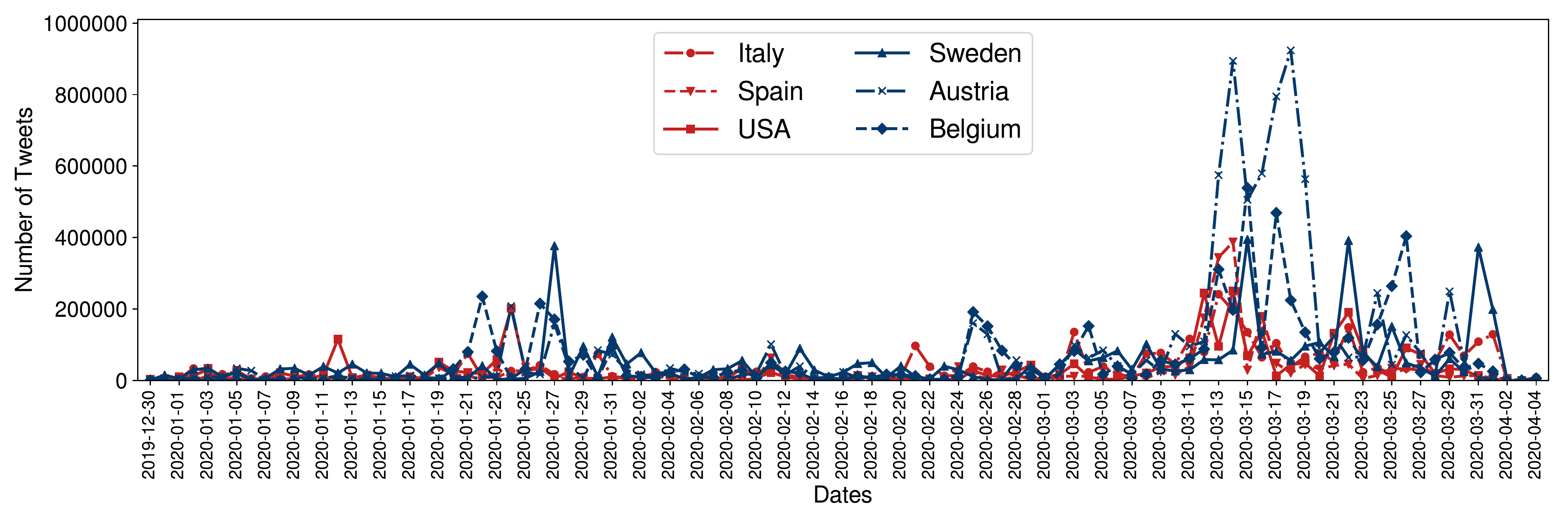}}
\vspace{-4mm}
    \caption{Temporal patterns in Trends and Tweets related to COVID-19. The United States, Spain, and Italy are annotated in red color while Sweden, Austria, and Belgium are annotated in blue color. Overall, the number of trends and the volume of tweets in the United States, Spain, and Italy are considerably low compared to the other three countries.}
    \label{fig:cstudy}
\vspace{-4mm}
\end{figure*}
\begin{enumerate*}[label=\protect\circled{\arabic*}, font=\small\bfseries]
\item Overall, as the number of cases, increased in a country, the number of tweets and trends increased accordingly. However, the relationship was not always linear. In most cases, the number of tweets decreased while the number of cases kept growing. 
\item A few countries (\ie Sweden and Austria) preemptively responded to the pandemic by actively discussing COVID-19 before the increase in the number of cases.
\item In some countries (\ie Austria), we observed a constant recurrence in tweets and trends, indicating consistency of interest on the subject. To take a deeper look at these observations, we present a case study below. 
\end{enumerate*}

\BfPara{Case Study} For the case study, we selected the top three countries from~\autoref{tab:one}, namely the United States, Spain, and Italy, and three other countries at random, namely Sweden, Austria, and Belgium. In the United States, the first COVID-19 case was reported on January 21, 2020, and by April 5, 2020, the number of cases exceeded 300K. Similarly, for Spain and Italy, the first case was reported on January 31, and the total number of cases increased to 132K and 18K by April 5, respectively. For Sweden, Austria, and Belgium, the first case was reported on February 4, February 24, and February 4, while the total cases increased to 6K, 12K, and 100K by April 5, respectively. For simplicity of analysis, we divide these countries into two sets where the set $\mathcal{S}_{1}$ consists of the United States, Spain, and Italy, and the set $\mathcal{S}_{2}$ consists of Sweden, Austria, and Belgium. Note that 1) all the first cases in $\mathcal{S}_{1}$ were reported earlier than the first cases reported in $\mathcal{S}_{2}$, and 2) by April 05, the total number of cases in $\mathcal{S}_{1}$ were much higher than the total number of cases in $\mathcal{S}_{2}$.

\begin{figure*}[!t]
\hfill
\begin{subfigure}[USA \label{fig:fee-tp}]{\includegraphics[width=5.5cm]{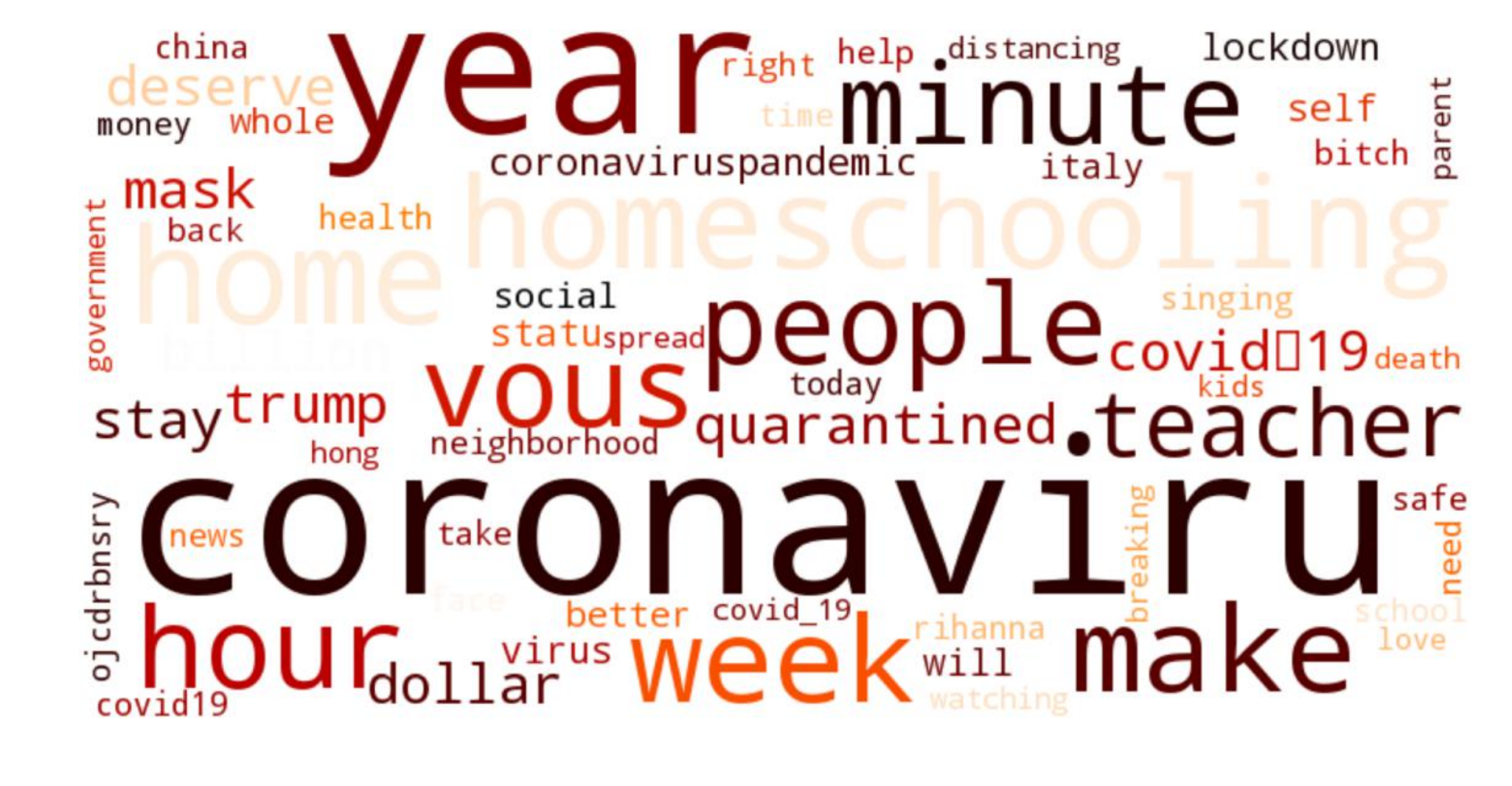}} 
\hfill
\end{subfigure}
\begin{subfigure}[Spain  \label{fig:fee-precision}]{\includegraphics[width=5.5cm]{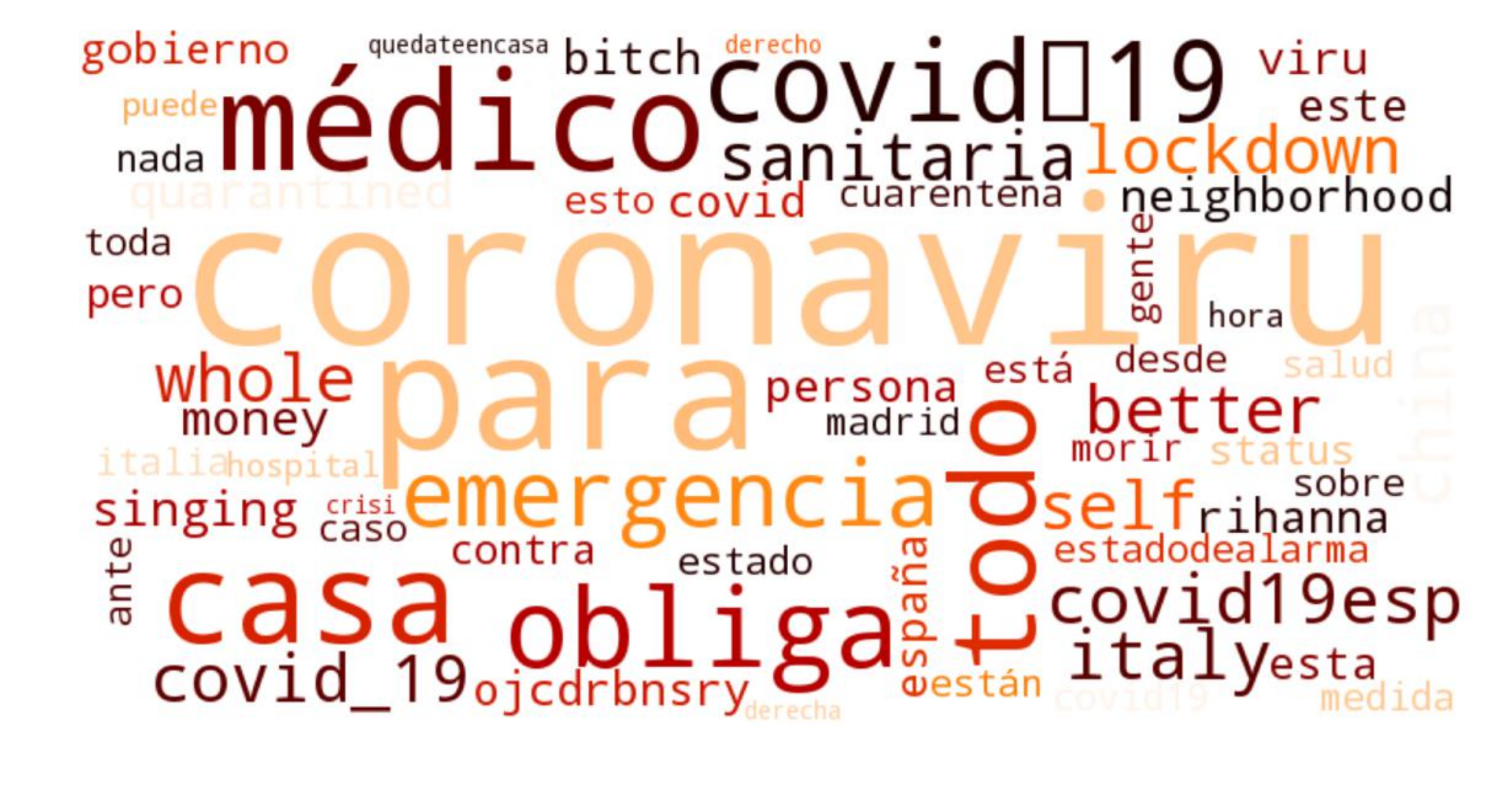}}
\hfill
\end{subfigure}
\begin{subfigure}[Italy  \label{fig:feediff}]{\includegraphics[width=5.5cm]{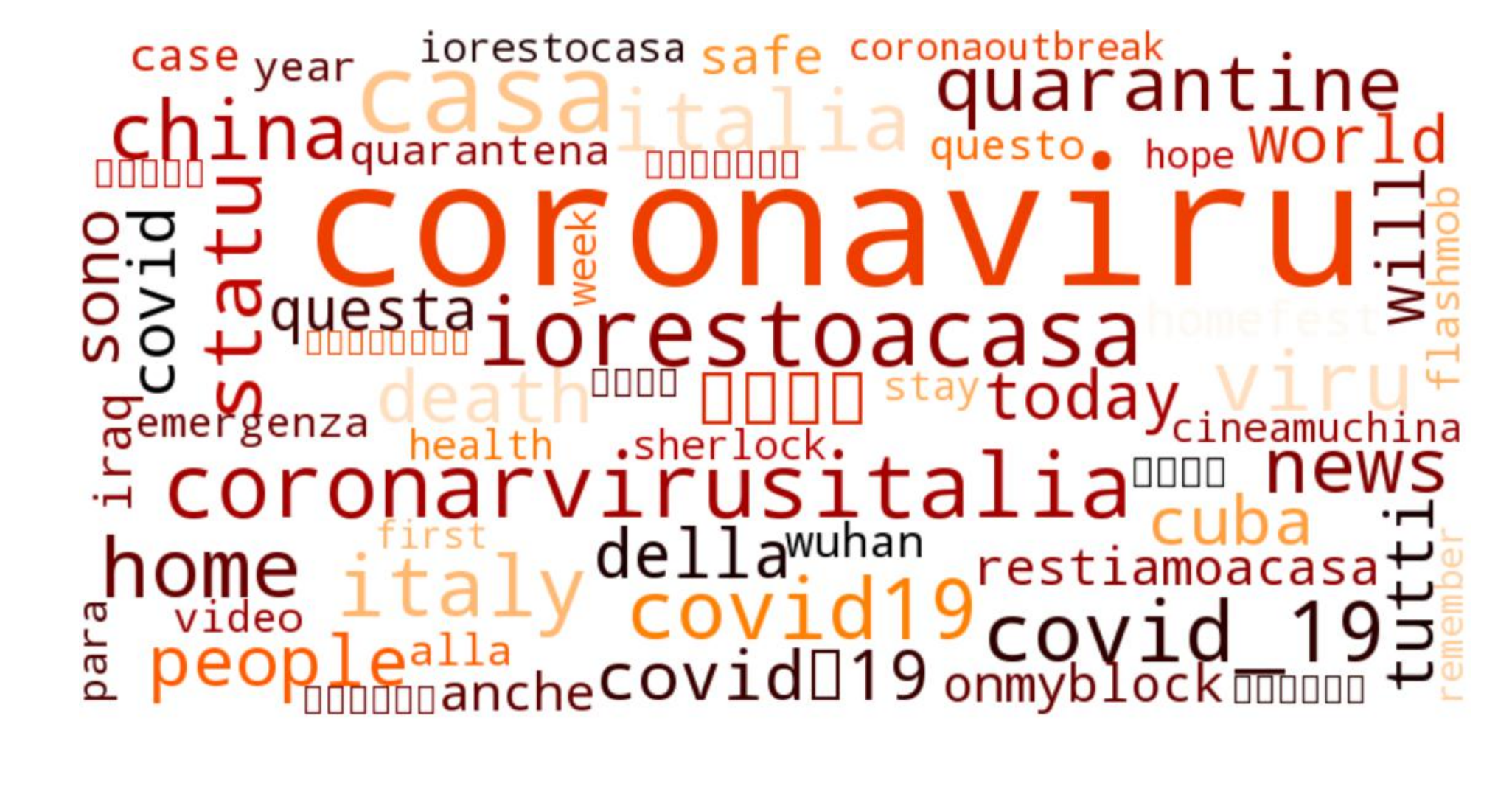}} 
\hfill
\end{subfigure}

\hfill
\begin{subfigure}[Belgium \label{fig:bgc}]{\includegraphics[width=5.5cm]{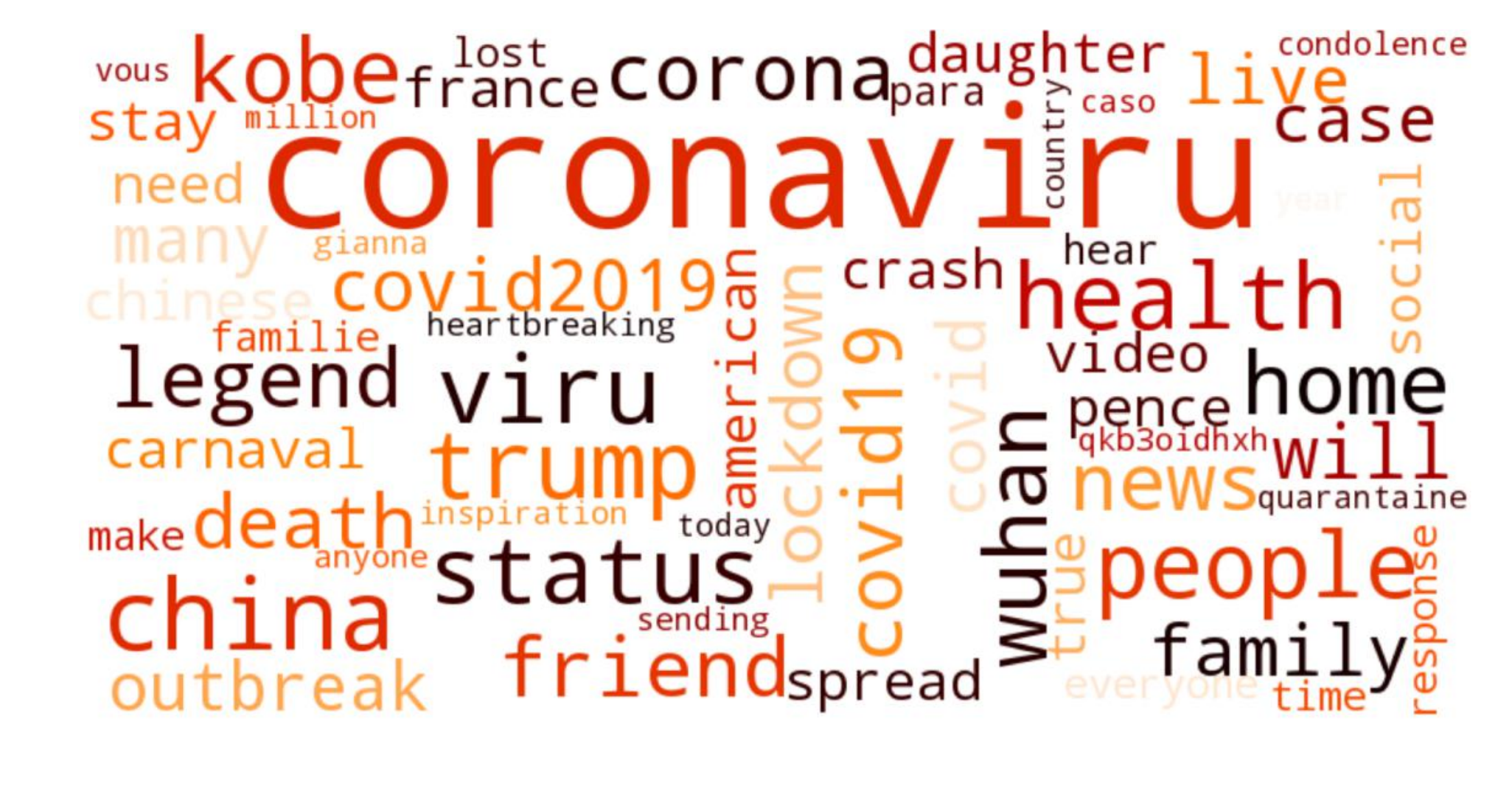}} 
\hfill
\end{subfigure}
\begin{subfigure}[Sweden  \label{fig:swcd}]{\includegraphics[width=5.5cm]{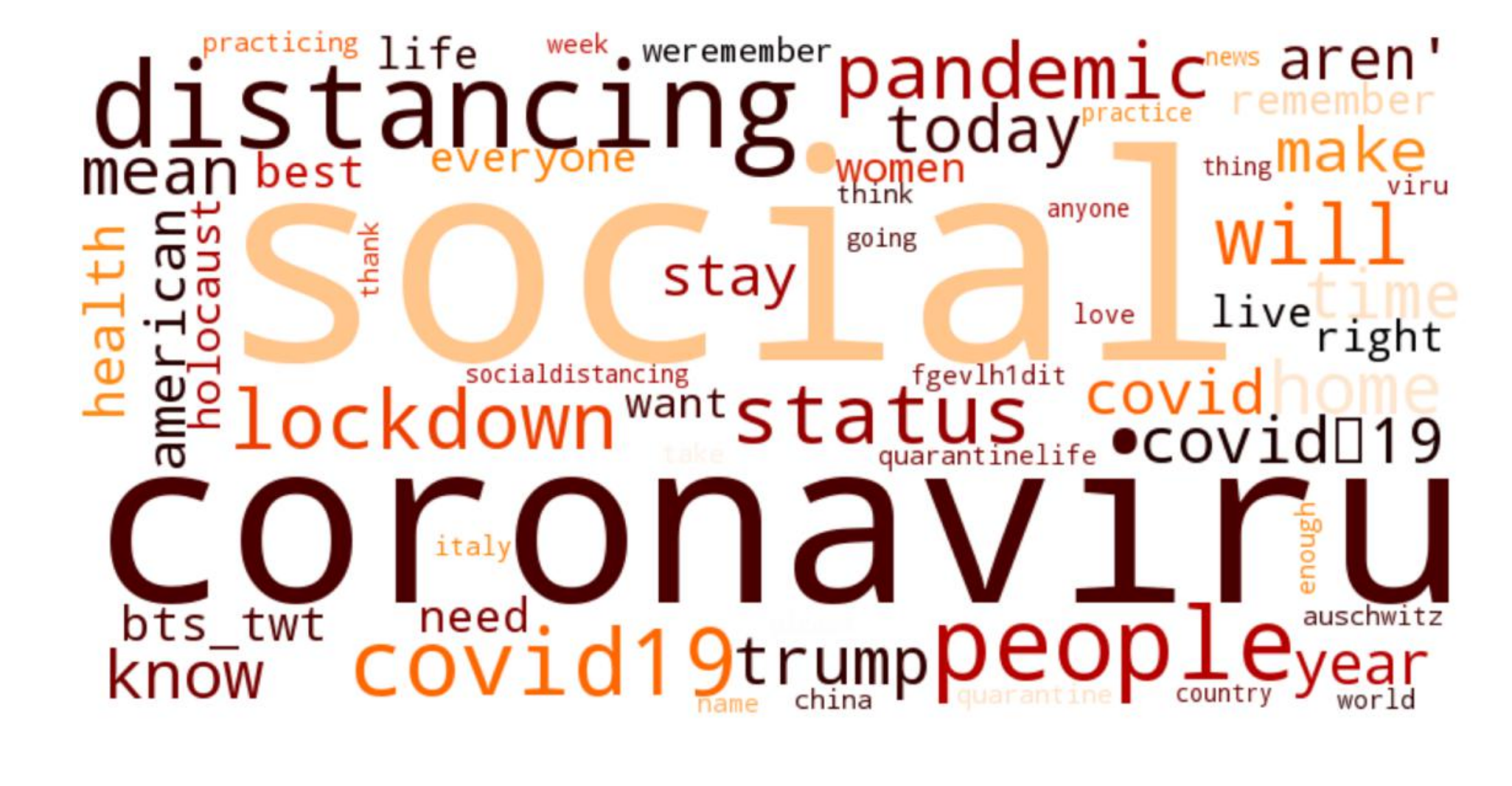}}
\hfill
\end{subfigure}
\begin{subfigure}[Austria  \label{fig:auscd}]{\includegraphics[width=5.5cm]{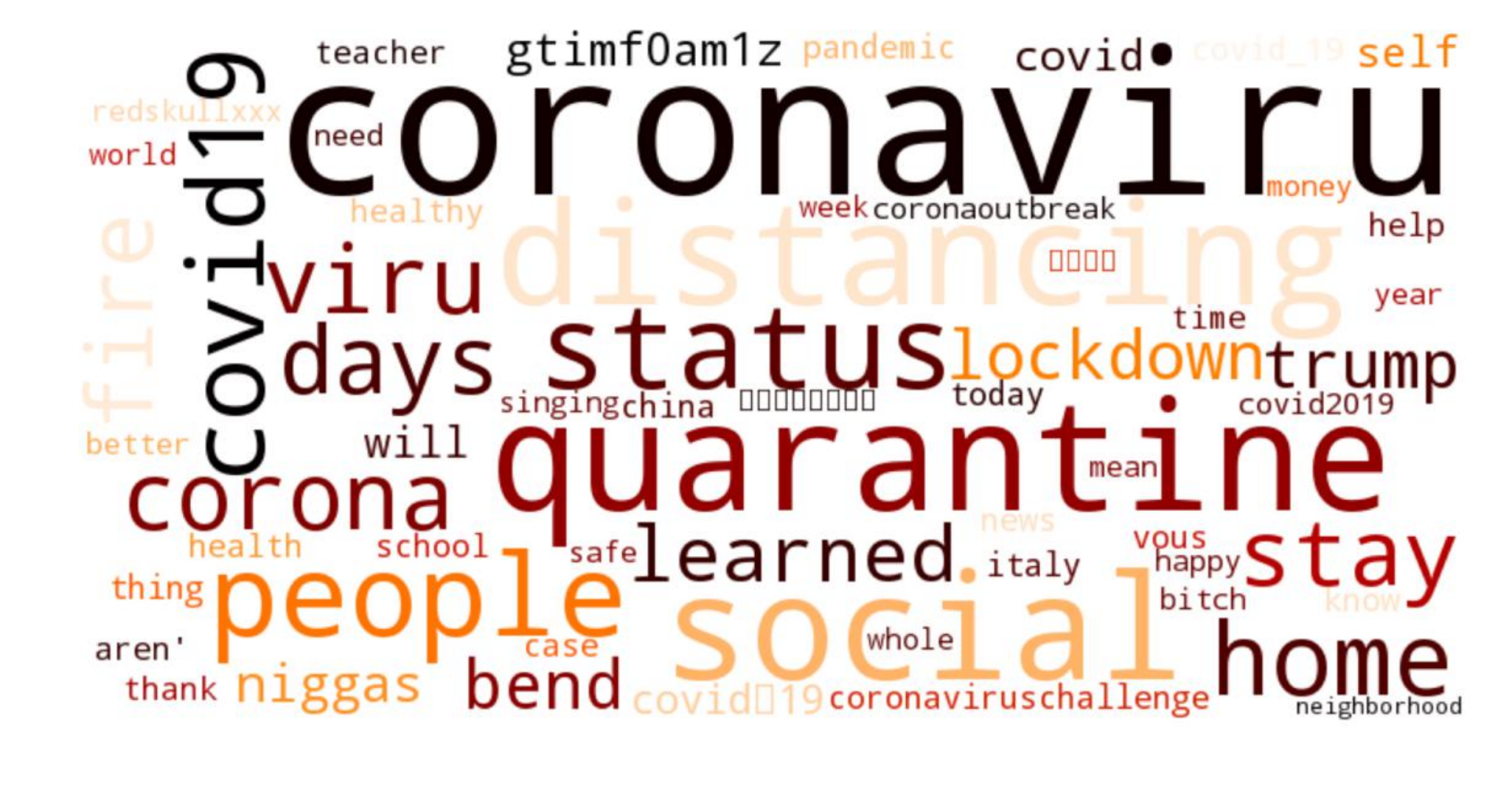}} 
\hfill
\end{subfigure}

\caption{Word clouds showing  prevalent topics in the COVID-19 tweets for $\mathcal{S}_{1}$ and $\mathcal{S}_{2}$. Overall, coronavirus is among the common terms in all countries. In Sweden and Austria, quarantine, social distancing, and lockdown are also prevalent. } 

\label{fig:cloud}
\end{figure*}

We analyzed the temporal behavior of Twitter trends in $\mathcal{S}_{1}$ and $\mathcal{S}_{2}$. For both sets, we apply~\autoref{algo:sk} to obtain the 1) the timeline of tweets in COVID-19 related trends, and 2) the total number of tweets about COVID-19. We separate tweets into two categories because the text in a tweet may or may not be associated with a COVID-19 trend. To understand this phenomenon, assume that an ongoing trend in a country is \#COVID--19. A user in that country tweets, ``Today, we have reported ten new cases of \#COVID--19.'' This tweet will appear in the \#COVID--19 trend and the trend will appear in our list of terms related to COVID-19. In contrast, if an ongoing trend in a country is \#FootballMatch and a user tweets ``\#FootballMatch has been canceled due to coronavirus,'' then the tweet will not appear in the COVID-19 related trends. However, the tweet will match in our list of terms related to COVID-19. The first example shows that COVID-19 is an actively discussed topic in a country since it appears among trends. If we sample all tweets related to COVID-19 related trends, we can estimate the significance of the topic in the country. However, this method may not capture complete information about tweets related to COVID-19, as demonstrated in the second example. Therefore, apart from acquiring a holistic view through COVID-19 trends, we also construct a complete picture by collecting all COVID-19 tweets from all trends, irrespective of the trend nature. The second method allows us to precisely determine the number of times the COVID-19 was discussed by people in a country. We use \autoref{algo:sk} to extract this information. We report our results in~\autoref{fig:cstudy} where~\autoref{fig:a} shows the total number of tweets related to COVID-19 trends and~\autoref{fig:b} shows the total number of COVID-19 tweets among all trends. The total number of tweets in both figures also include the number of retweets. 

\BfPara{Key Takeaways} Our results show that countries in $\mathcal{S}_{2}$ generated COVID-19 related tweets and trends before countries in $\mathcal{S}_{1}$, indicating a preemptive attempt to cause pandemic awareness. In the entire evaluation timeline, all countries in $\mathcal{S}_{2}$ generated more COVID-19 tweets than countries in $\mathcal{S}_{1}$. We also observed spikes in~\autoref{fig:cstudy}, showcasing a surge in the number of tweets and trends. In all noticeable spikes, $\mathcal{S}_{2}$ clearly dominated $\mathcal{S}_{1}$, indicating a higher user engagement towards COVID-19. Among all countries, Switzerland generated the highest number of COVID-19 tweets ($\approx$7.3 Million) and the highest number of COVID-19 trends (272). The inner plot in~\autoref{fig:a} shows that number of daily trends in $\mathcal{S}_{2}$ were considerably higher than $\mathcal{S}_{1}$. Notably, in Belgium, 15 COVID-19 trends were generated on March 19, 2020. These results show that countries in $\mathcal{S}_{2}$ effectively utilized Twitter to propagate information among users and prepare them for the pandemic.

{
\renewcommand{\arraystretch}{1.3}
\begin{table}[t]
\centering
\caption{Top 10 most common words in the text corpus of each country. Note that the three most common words in Sweden are Social, Distancing, and Coronavirus. }
\scalebox{0.9}{\begin{tabular}{ll}
\hline\hline
\textbf{Country} & \textbf{10 Most Common Topics} \\ \hline
USA              & Coronavirus, Year, Home, Homeschooling \\ &
Make, Week, Vous, People, Minute, Hour
            \\ \hline
Spain           & Coronavirus, Para, Médico, Casa, Todo, Covid-19 \\ &
Obliga, Emergencia, Covid19esp, Italy \\ \hline
Italy            & Coronavirus, Casa, Iorestoacasa, Coronarvirusitalia, \\ & Covid_19  
Italy, Covid19, Siru, Statu, Home
 \\ \hline
Belgium          &  Coronaviru, China, Virus, Trump, Status, \\ &
Health, People, Kobe, Home, Wuhan              \\ \hline
Austria         & Coronavirus, Quarantine, Social, Distancing, Status, \\ &
People, Covid19, Home, Corona, Virus \\ \hline
Sweden         & Social, Distancing, Coronavirus, People \\ &
Covid19, Status, Lockdown, Will, Pandemic, Home \\ \hline\hline
\end{tabular}} \label{tab:two}
\end{table}
}

\subsection{Topic Modeling} \label{sec:tm}
In our second experiment, we take a closer look at the textual information in the tweet corpus to make useful inferences about prevalent topics in those tweets. To motivate a common case, we limit our analysis to $\mathcal{S}_{1}$ and $\mathcal{S}_{1}$, and retrieve their tweet corpus from~\autoref{algo:sk}. 

\begin{figure*}[t]
    \centering
    
\subfigure[Sentiments on Social Distancing]{\label{fig:sd}\includegraphics[width=1.05\textwidth]{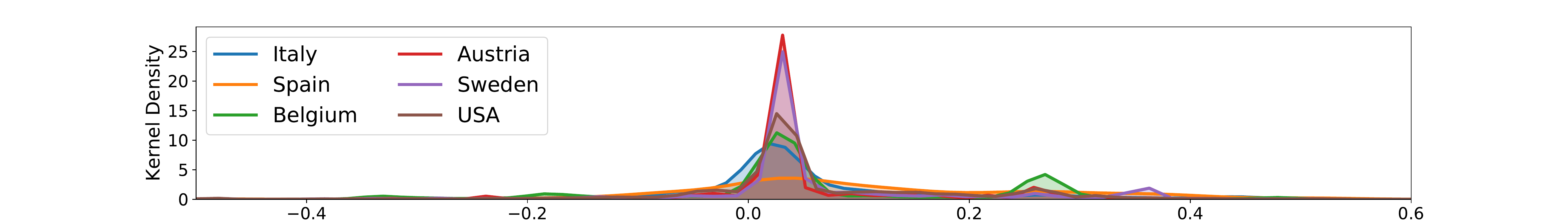}}

\subfigure[Sentiments on Quarantine]{\label{fig:qr}\includegraphics[width=1.05\textwidth]{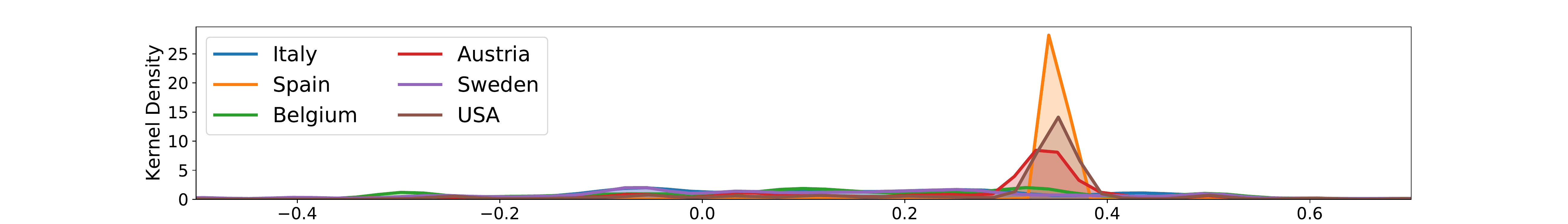}}

\subfigure[Sentiments on LockDown]{\label{fig:ld}\includegraphics[width=1.05\textwidth]{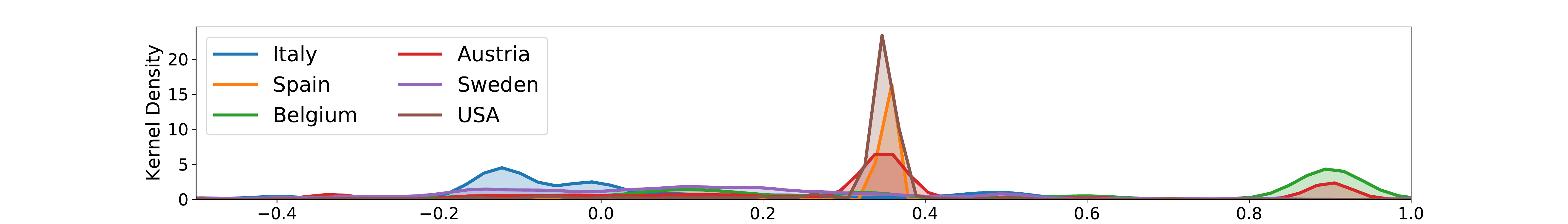}}
\vspace{-3mm}
    \caption{Sentiment analysis of users towards social distancing, quarantine, and lockdown. The x-axis shows the sentiment score between the range of -1 and +1. The y-axis shows the kernel density estimation that captures the data distribution shape. Overall, the general sentiment in each class closely aligns across all countries. For social distancing, the sentiment is close to neutral. For quarantine and social distancing, the sentiment is more distributed towards the positive side.}
    \label{fig:sa}
\vspace{-3mm}
\end{figure*}

To study prevalent topics among COVID-19 tweets, we combined those tweets in a single text corpus for each country. We then tokenized the text corpus, removed the stop words, and calculated the frequency count over the resulting text. Finally, we assigned weights to all the topics and sorted them in descending order. In~\autoref{fig:cloud}, we show word clouds for each country, providing an intuitive overview of the most commonly used terms in COVID-19 tweets. \autoref{fig:cloud} shows that generally, ``coronavirus'' was the most common term across all countries. Noticeably, in Sweden and Austria, ``social,'' ``distancing'', ``quarantine,'' ``lockdown,'' ``stay,'' and ``home'' were the more dominant compared to other countries. Since it is possible that the two terms ``social'' and ``distancing'' may appear in different contexts across tweets, therefore we performed the same experiment while incorporating bigrams model. The bigram model approximates the probability of a word by conditioning over a preceding word. As such, if ``social'' and ``distancing'' are collocated, then they would naturally appear in the model. We report the results of the bigram model in~\autoref{fig:cloud2}. Our results confirmed that ``social distancing'' and ``stay home'' were indeed dominant terms in Sweden and Austria.

Additionally, in \autoref{tab:two}, we report the ten most common terms that appeared in our topic modeling. \autoref{tab:two} shows that the trending topics significantly varied in each country. The common term among all countries was ``Coronavirus,'' followed by ``Covid19.''  Moreover, in all countries except Sweden, ``Coronavirus'' was the most common term. In Sweden, the top two terms were ``Social'' and ``Distancing,'' indicating that the Twitter users in Sweden significantly emphasized on the preventive measures. Combined, the number of COVID-19 topics in $\mathcal{S}_{2}$ were greater than $\mathcal{S}_{1}$. Although, considering the total number of COVID-19 cases and deaths in $\mathcal{S}_{1}$, we expected the outcome to be the opposite.





\subsection{Sentiment Analysis} \label{sec:sentiment}
In our third experiment, we analyze the user sentiments towards the COVID-19 related preventive measures. Towards that, first, we isolated tweets containing terms ``social distancing,'' ``quarantine,'' and ``lockdown.'' We distributed those tweets in three separate classes. Additionally, we also incorporated terms that closely resembled a specified class. For instance, ``curfew'' closely relates to the class ``lockdown,'' while ``self isolation'' relates to the class ``quarantine.'' We manually annotated such similar terms and incorporated them into the corresponding classes. 

For sentiment analysis, we used the ``TextBlob'' library in Python that provides various useful language processing operations, including speech tagging, text tokenization, sentiment analysis, and sentiment classification. The ``TextBlob'' library assigns a score in the range of -1 to 1 to each tweet in the class. We eliminated tweets with a neutral score of ``0'' to focus purely on tweets with a positive or negative sentiment. Additionally, we applied the kernel density function to aggregate tweets with the same sentiment score and observed the distribution shape of each class.    
We report our results in~\autoref{fig:sa}. Our results show that for social distancing and quarantine; generally, the sentiment across all countries was within the same margin. For social distancing, almost all countries had a close to neutral sentiment, as indicated by a spike around 0.1~\autoref{fig:sd}. However, we also observed a small spike towards the positive sentiment in  Belgium and Sweden. Similarly, for the quarantine class, we noticed a spike of around 0.3 for all countries, indicating a more positive response. For the lockdown class, we observed a relatively higher sentiment variation. In Italy, the sentiment was distributed towards the negative side, with a spike around -0.1~\autoref{fig:ld}. However, in Austria and Belgium, the sentiment was allocated towards the positive side with peaks around 0.9.
In summary, our data show that the general response to social distancing and quarantine was similar across all countries. However, for lockdown, we observed a variation in response with Italy's inclination towards a negative sentiment. In summary, the general sentiment on social distancing and quarantine, across all countries, converged to a similar score in the density distribution. This observation reflects a sense of uniformity in expression for all countries. However, for lockdown, the variation in score indicates a divergence in expression towards it. This could be a result of societal pressures operating in those countries which we could not capture in our dataset. Perhaps a more precise coupling of sentiment with the increasing number of cases will provide reasoning for the sentiment divergence. In the future, we plan to explore this direction and get more meaningful results to support the observation.

\section{Discussion and Conclusion} \label{sec:conc}
As discussed in~\tsref{sec:intro}, social media platforms can be useful in characterizing public opinion in a geographical locality. Additionally, these platforms can also be used to monitor the effects of information propagation by pairing the information flow with a desirable outcome. In this paper, we contextualize this methodology to study the relationship between information dissemination the COVID-19 pandemic spread. Our model puts ``lower spread'' as the desirable outcome and ``high volumes of trends and tweets'' as the indicators of effective information dissemination. To that end, we developed a large-scale data collection system to collect historical tweets from the top 20 most affected countries by COVID-19. We perform measurements and modelling on our data to study various data attributes including the temporal evolution of trends, the most recurring COVID-19 related topics, and the user sentiment towards preventive measures. 

\begin{figure*}[t]
\hfill
\begin{subfigure}[USA \label{fig:fee-tp2}]{\includegraphics[width=5.5cm]{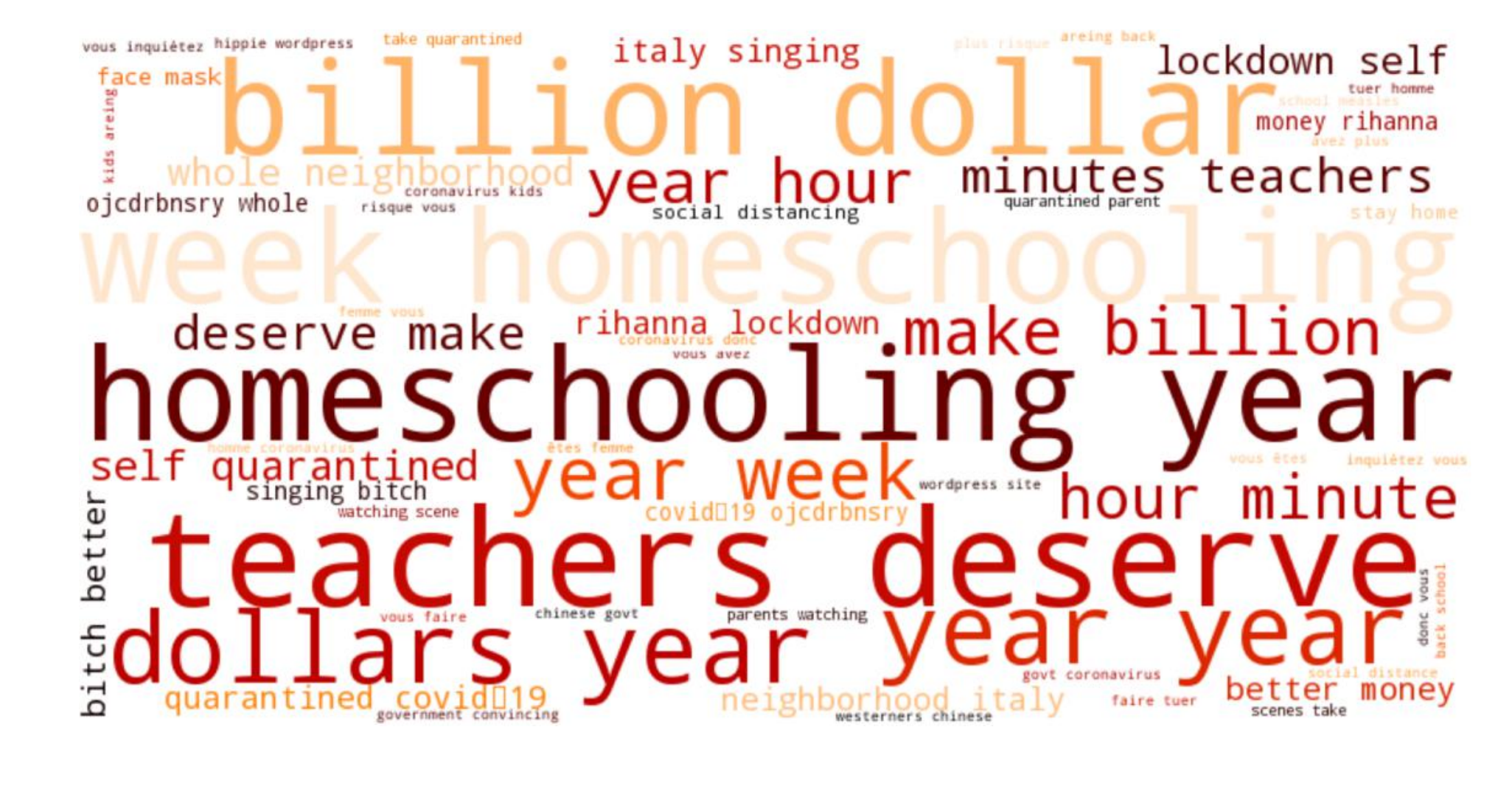}} 
\hfill
\end{subfigure}
\begin{subfigure}[Spain  \label{fig:fee-precision2}]{\includegraphics[width=5.5cm]{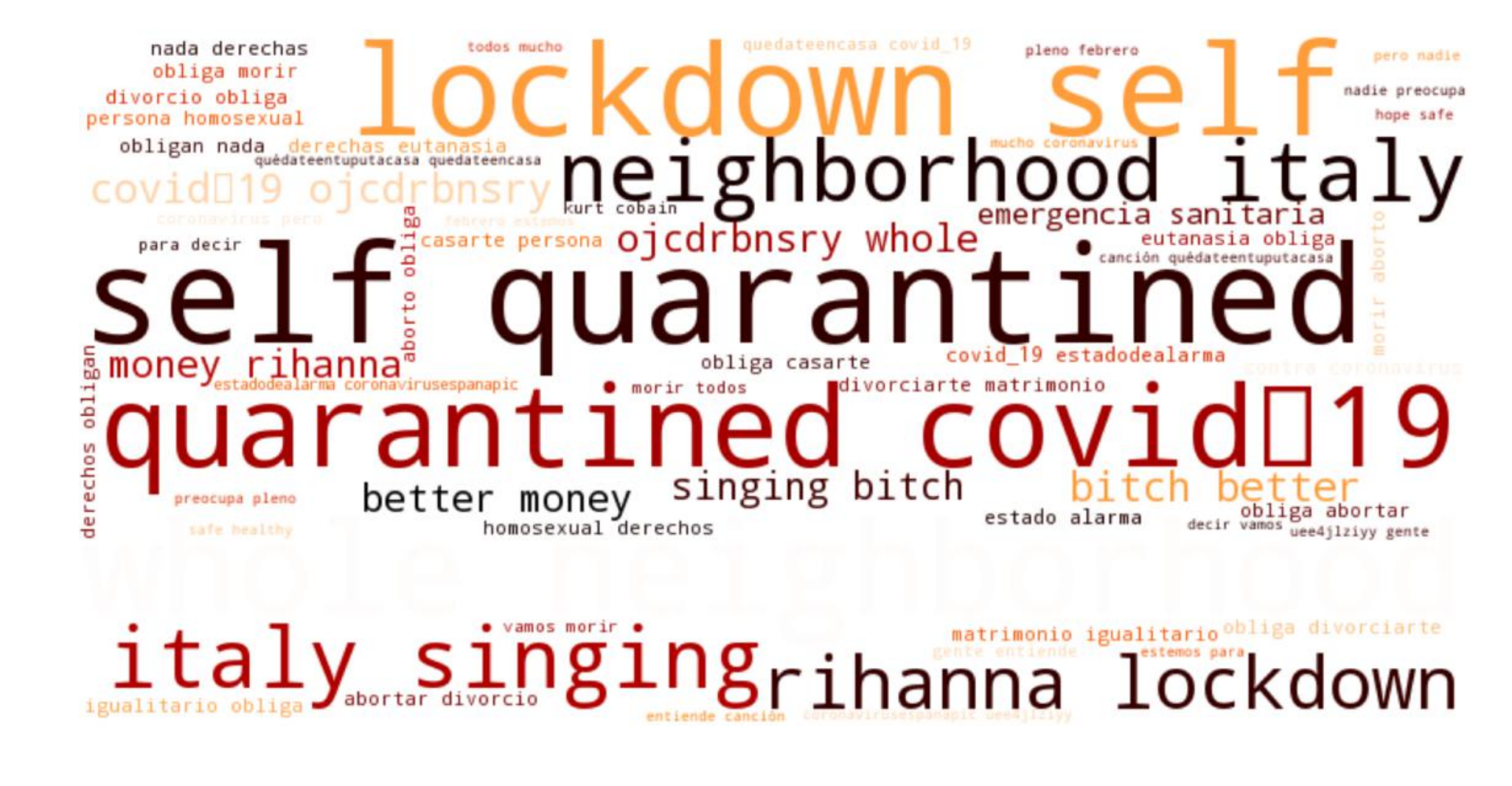}}
\hfill
\end{subfigure}
\begin{subfigure}[Italy  \label{fig:feediff2}]{\includegraphics[width=5.5cm]{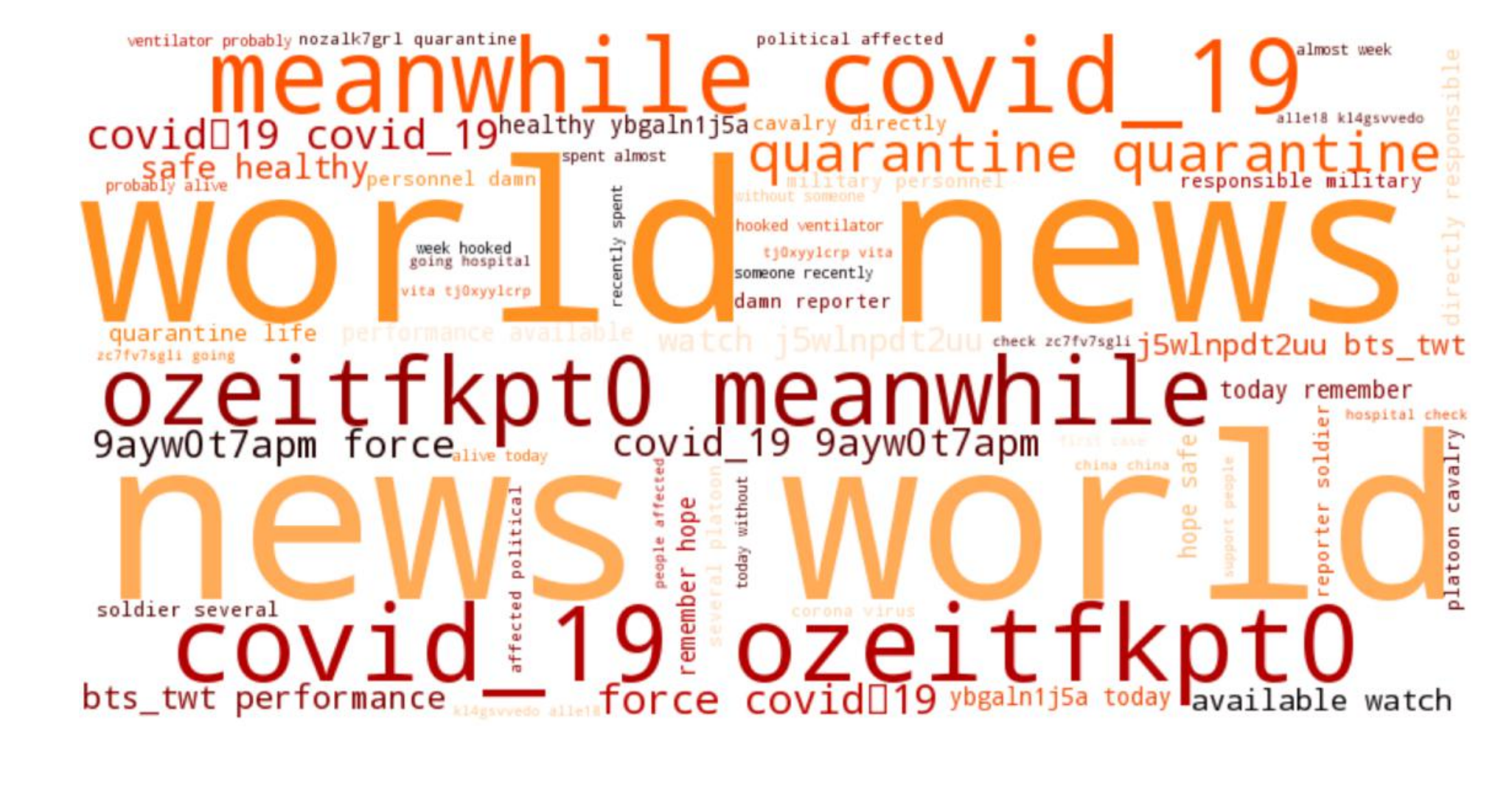}} 
\hfill
\end{subfigure}

\hfill
\begin{subfigure}[Belgium \label{fig:bgc2}]{\includegraphics[width=5.5cm]{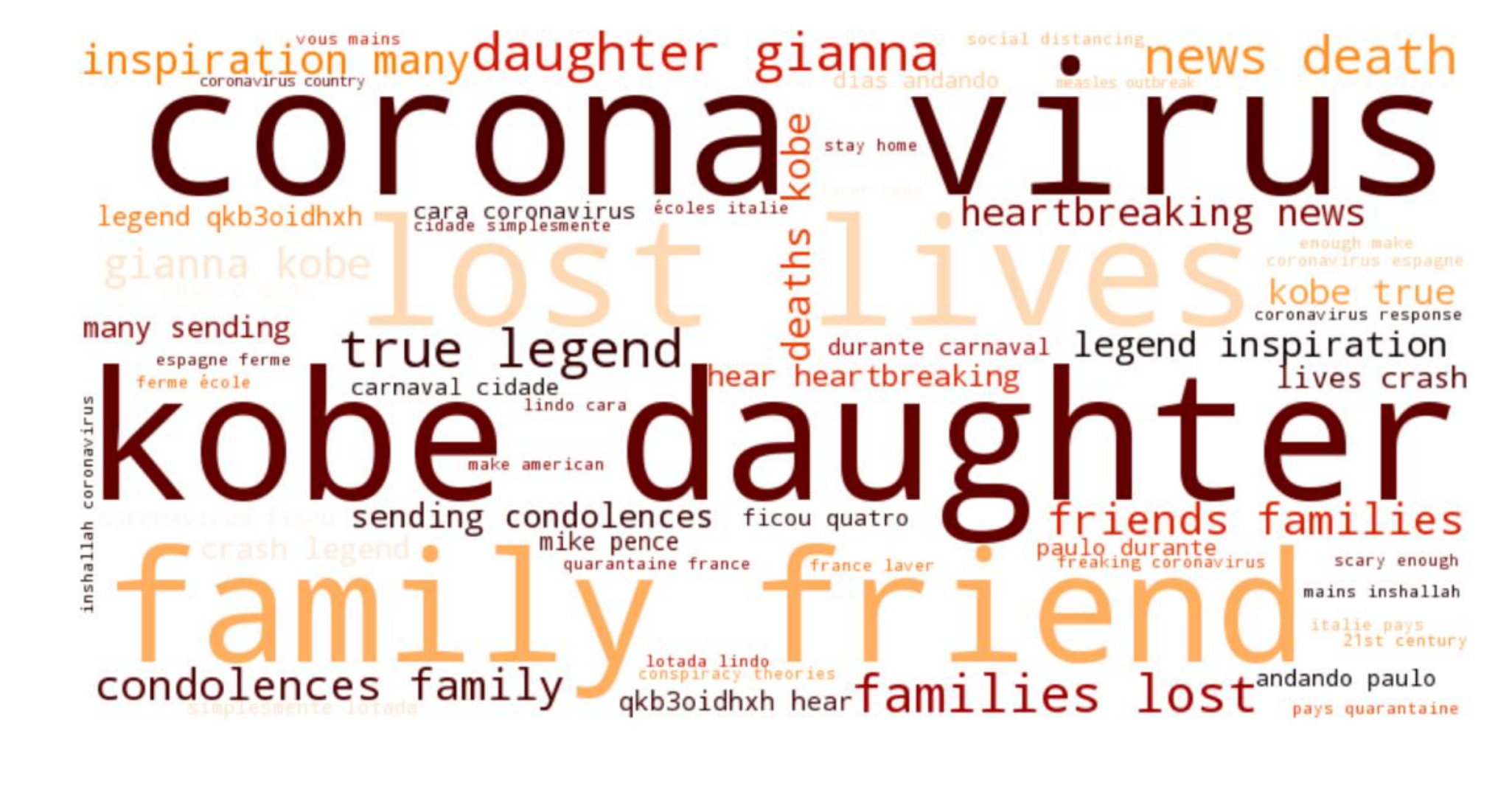}} 
\hfill
\end{subfigure}
\begin{subfigure}[Sweden  \label{fig:swcd2}]{\includegraphics[width=5.5cm]{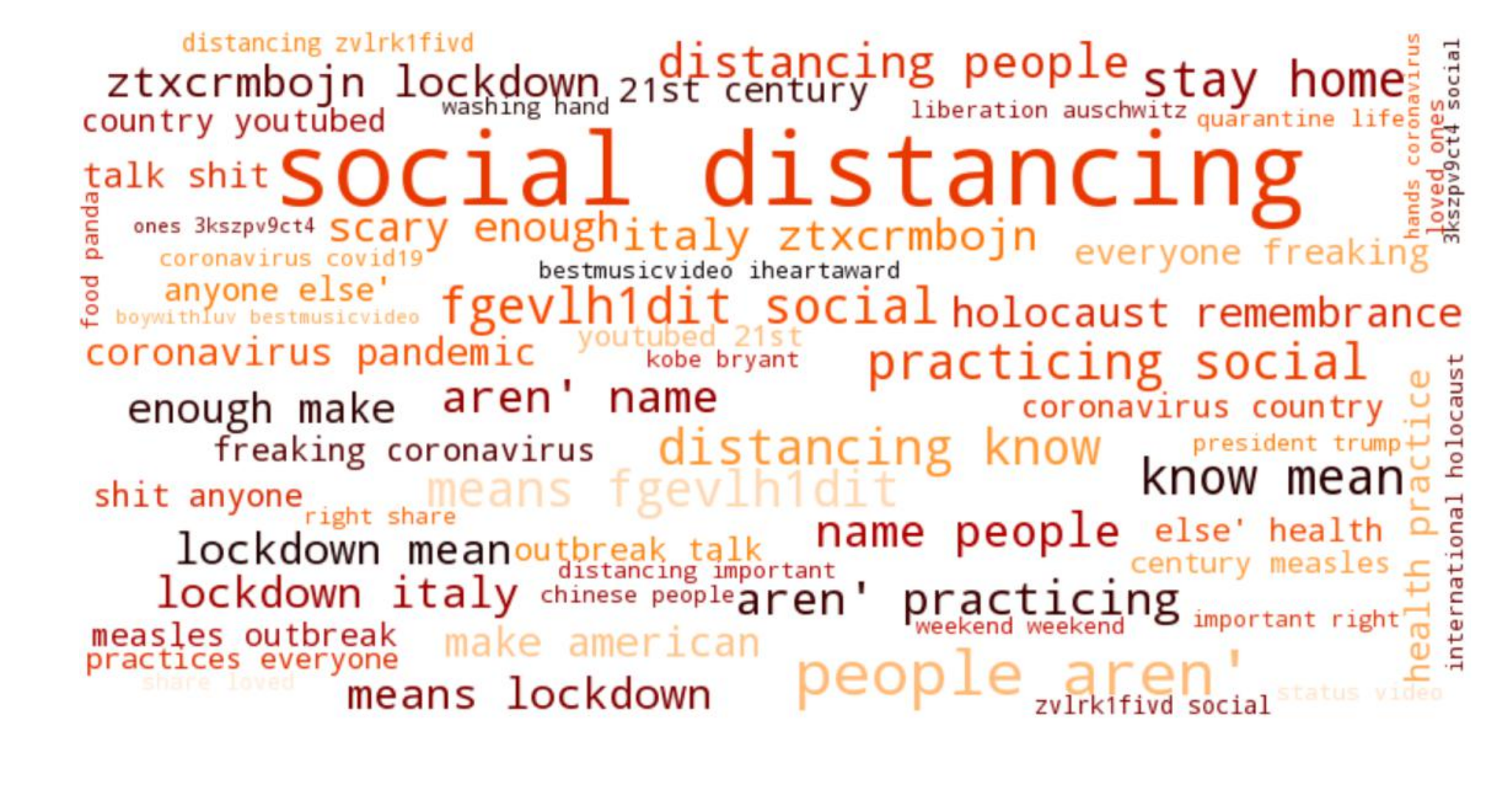}}
\hfill
\end{subfigure}
\begin{subfigure}[Austria  \label{fig:auscd2}]{\includegraphics[width=5.5cm]{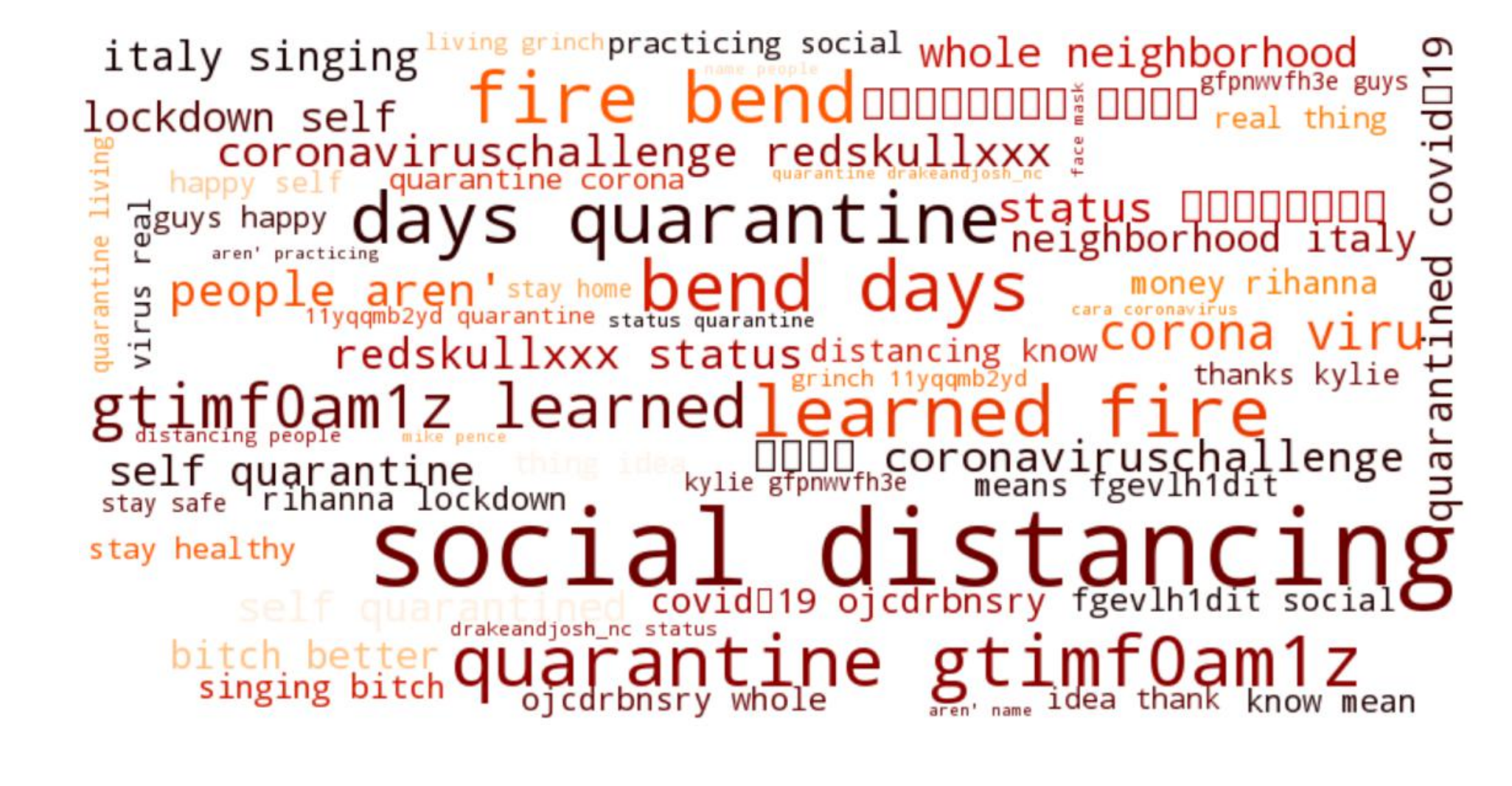}} 
\hfill
\end{subfigure}

\caption{Word clouds for  $\mathcal{S}_{1}$ and $\mathcal{S}_{2}$ after bigram analysis. Notice that for Sweden and Austria, ``Social Distancing'' was the most dominant term. In contrast, for Spain, self quarantined was the most dominant term.  } 

\label{fig:cloud2}
\vspace{-4mm}
\end{figure*}

Our results show that countries with a lower pandemic spread mostly generated a higher volume of COVID-19 related trends and tweets (\autoref{tab:one}, \autoref{fig:cstudy}). A closer look at the nature of tweets further revealed that the countries with a lower pandemic spread emphasized more on the COVID-19 preventive measures (\autoref{fig:cloud}, \autoref{fig:cloud2}). Moreover, we also noticed a variation in sentiment towards the lockdown policy that was implemented to control the spread.

In addition to making standalone contributions through a novel dataset and useful observations, our study also provides meaningful answers to the questions raised in~\tsref{sec:intro}. First, we indeed noticed variations in the response of different countries to the COVID-19 pandemic as shown by the 1) volume of trends and tweets and their timeline, 2) recurring topics discussed in those tweets, and 3) sentiments towards preventive measures. Second, we also observed indications to support that awareness through Twitter contributed in influencing the pandemic spread. For that purpose, we outlined a case study to showcase that users in the highly affected countries displayed lower Twitter engagement compared to the lesser affected countries. Please note that this is not a conclusive statement to suggest that Twitter usage was the dominant factor in influencing the pandemic spread. However, our data and analysis indicate that Twitter can be useful for this purpose, and therefore noteworthy.

\BfPara{Future Work} At the time of conducting this study, we did not find a study that precisely analyzed the relationship between Twitter and the spread of COVID-19. However, our methodology is inspired from some notable studies that examined the usefulness of Twitter in characterizing the user behavior at scale. We have mentioned them in~\tsref{sec:intro}. Concurrent to work, we have seen a study that analyzed the emergence of Sinophobic behavior due to COVID-19~\cite{SchildLBSZZ20}. However, our work investigates an entirely different relationship between Twitter and COVID-19. 

Finally, we believe that our dataset has useful information beyond what is presented in this paper. Keeping it in view, as well as the urgency to extend research on this topic, we will soon open-source our dataset to foster future work.  

\bibliographystyle{aaai}
\bibliography{ref}

\begin{thebibliography}{}

\bibitem[\protect\citeauthoryear{An \bgroup et al\mbox.\egroup
  }{2018}]{AnYLDZL18}
An, L.; Yu, C.; Lin, X.; Du, T.; Zhou, L.; and Li, G.
\newblock 2018.
\newblock Topical evolution patterns and temporal trends of microblogs on
  public health emergencies: An exploratory study of ebola on twitter and
  weibo.
\newblock {\em Online Information Review} 42(6):821--846.

\bibitem[\protect\citeauthoryear{ArchiveCommunity}{2020}]{wayback_machine20}
ArchiveCommunity.
\newblock 2020.
\newblock Wayback machine apis.

\bibitem[\protect\citeauthoryear{Bin~Tareaf \bgroup et al\mbox.\egroup
  }{2018}]{TareafBHKKMC18}
Bin~Tareaf, R.; Berger, P.; Hennig, P.; Koall, S.; Kohstall, J.; and Meinel, C.
\newblock 2018.
\newblock Information propagation speed and patterns in social networks: A case
  study analysis of german tweets.
\newblock {\em Journal of Computers} 13:761--770.

\bibitem[\protect\citeauthoryear{Brena \bgroup et al\mbox.\egroup
  }{2019}]{Brena0CGPR19}
Brena, G.; Brambilla, M.; Ceri, S.; Giovanni, M.~D.; Pierri, F.; and Ramponi,
  G.
\newblock 2019.
\newblock News sharing user behaviour on twitter: {A} comprehensive data
  collection of news articles and social interactions.
\newblock In {\em Proceedings of the Thirteenth International Conference on Web
  and Social Media, {ICWSM} 2019, Munich, Germany, June 11-14, 2019},
  592--597.

\bibitem[\protect\citeauthoryear{Broniec \bgroup et al\mbox.\egroup
  }{2020}]{BroniecARG20}
Broniec, W.; An, S.; Rugaber, S.; and Goel, A.~K.
\newblock 2020.
\newblock Using {VERA} to explain the impact of social distancing on the spread
  of {COVID-19}.
\newblock {\em CoRR} abs/2003.13762.

\bibitem[\protect\citeauthoryear{COVID-19}{2020}]{COVID19Timeline}
COVID-19.
\newblock 2020.
\newblock Covid-19 pandemic by country and territory.

\bibitem[\protect\citeauthoryear{Fischer{-}Pre{\ss}ler, Schwemmer, and
  Fischbach}{2019}]{Fischer-Pressler19}
Fischer{-}Pre{\ss}ler, D.; Schwemmer, C.; and Fischbach, K.
\newblock 2019.
\newblock Collective sense-making in times of crisis: Connecting terror
  management theory with twitter user reactions to the berlin terrorist attack.
\newblock {\em Comput. Hum. Behav.} 100:138--151.

\bibitem[\protect\citeauthoryear{Inoue and Todo}{2020}]{InoueT20}
Inoue, H., and Todo, Y.
\newblock 2020.
\newblock The propagation of the economic impact through supply chains: The
  case of a mega-city lockdown against the spread of {COVID-19}.
\newblock {\em CoRR} abs/2003.14002.

\bibitem[\protect\citeauthoryear{Katella}{2020}]{katella_2020}
Katella, K.
\newblock 2020.
\newblock Our new covid-19 vocabulary-what does it all mean?

\bibitem[\protect\citeauthoryear{Keymanesh \bgroup et al\mbox.\egroup
  }{2019}]{KeymaneshGBBCP19}
Keymanesh, M.; Gurukar, S.; Boettner, B.; Browning, C.~R.; Calder, C.~A.; and
  Parthasarathy, S.
\newblock 2019.
\newblock Twitter watch: Leveraging social media to monitor and predict
  collective-efficacy of neighborhoods.
\newblock {\em CoRR} abs/1911.06359.

\bibitem[\protect\citeauthoryear{Le, Shafiq, and Srinivasan}{2017}]{LeSS17}
Le, H.~T.; Shafiq, Z.; and Srinivasan, P.
\newblock 2017.
\newblock Scalable news slant measurement using twitter.
\newblock In {\em Proceedings of the Eleventh International Conference on Web
  and Social Media, {ICWSM} 2017, Montr{\'{e}}al, Qu{\'{e}}bec, Canada, May
  15-18, 2017},  584--587.

\bibitem[\protect\citeauthoryear{Mottl}{2019}]{mottl_2019}
Mottl.
\newblock 2019.
\newblock Crawloldtweets.

\bibitem[\protect\citeauthoryear{Pierce and Center}{2020}]{tmc_2020}
Pierce, S., and Center, T.~M.
\newblock 2020.
\newblock Covid-19 crisis catalog: A glossary of terms.

\bibitem[\protect\citeauthoryear{Pratikakis}{2018}]{Pratikakis18}
Pratikakis, P.
\newblock 2018.
\newblock twawler: {A} lightweight twitter crawler.
\newblock {\em CoRR} abs/1804.07748.

\bibitem[\protect\citeauthoryear{Robson}{2020}]{Robson20a}
Robson, B.
\newblock 2020.
\newblock Computers and viral diseases. preliminary bioinformatics studies on
  the design of a synthetic vaccine and a preventative peptidomimetic
  antagonist against the sars-cov-2 (2019-ncov, {COVID-19)} coronavirus.
\newblock {\em Comp. in Bio. and Med.} 119:103670.

\bibitem[\protect\citeauthoryear{Schild \bgroup et al\mbox.\egroup
  }{2020}]{SchildLBSZZ20}
Schild, L.; Ling, C.; Blackburn, J.; Stringhini, G.; Zhang, Y.; and Zannettou,
  S.
\newblock 2020.
\newblock "go eat a bat, chang!": An early look on the emergence of sinophobic
  behavior on web communities in the face of {COVID-19}.
\newblock {\em CoRR} abs/2004.04046.

\bibitem[\protect\citeauthoryear{TrendoGateCommunity}{2020}]{trendogate20}
TrendoGateCommunity.
\newblock 2020.
\newblock Twitter trends archive trends everywhere anytime.

\bibitem[\protect\citeauthoryear{Tulasi \bgroup et al\mbox.\egroup
  }{2019}]{TulasiGGSMBK19}
Tulasi, A.; Gupta, K.; Gurjar, O.; Buggana, S.~S.; Mehan, P.; Buduru, A.~B.;
  and Kumaraguru, P.
\newblock 2019.
\newblock Catching up with trends: The changing landscape of political
  discussions on twitter in 2014 and 2019.
\newblock {\em CoRR} abs/1909.07144.

\bibitem[\protect\citeauthoryear{Valkanas, Saravanou, and
  Gunopulos}{2014}]{ValkanasSG14}
Valkanas, G.; Saravanou, A.; and Gunopulos, D.
\newblock 2014.
\newblock A faceted crawler for the twitter service.
\newblock In {\em Web Information Systems Engineering - {WISE} 2014 - 15th
  International Conference, Thessaloniki, Greece, October 12-14, 2014,
  Proceedings, Part {II}},  178--188.

\bibitem[\protect\citeauthoryear{Wells \bgroup et al\mbox.\egroup
  }{2020}]{WellsSLPPY20}
Wells, C.; Shah, D.~V.; Lukito, J.; Pelled, A.; Pevehouse, J. C.~W.; and Yang,
  J.
\newblock 2020.
\newblock Trump, twitter, and news media responsiveness: {A} media systems
  approach.
\newblock {\em New Media {\&} Society} 22(4).

\end{thebibliography}
\section{Appendices}\label{sec:append}

\subsection{Kernel Density Estimation} \label{sec:kd}

Kernel Density Estimator (KDE) is a renowned probability density function that is used to solve the data smoothing problem for a finite dataset. Typically, this is done by graphing the density of the dataset in its domain. The formal definition of KDE is given by the following function.

$$
\widehat{p}_{n}(x)=\frac{1}{n h} \sum_{i=1}^{n} K\left(\frac{X_{i}-x}{h}\right)
$$

In the function above, \textit{K(x)} is the smooth and symmetric \textit{kernel function} (Gaussian in our case), and \textit{h} (where \textit{h} $>$ 0) is the smoothing bandwidth. KDE calculates summation, after each data-point is smoothed into small density bumps.  

\subsection{Bigram Model Results} \label{sec:bm}

In \autoref{fig:cloud2}, we have generated the Bigram Model for countries discussed in our case studies. A Bigram Model looks one word into the past and predicts the next word. Building onto this, the \autoref{fig:cloud2} shows that which two words, together, are most likely to appear for each country in $\mathcal{S}_{1}$ and $\mathcal{S}_{2}$. 


Referring to \autoref{fig:cloud2}, we observe word clouds for  $\mathcal{S}_{1}$ and $\mathcal{S}_{2}$ after bigram analysis. The most common term in Sweden and Austria was "social distancing", and similarly, "corona virus" was dominant in Belgium. In the USA, "billion dollar" was dominant along with "homeschooling year", in Spain's word cloud, the most common term was "self quarantined", and in Italy, the common term was "world news".

\end{document}